\documentclass[aps,pre,twocolumn,groupedaddress]{revtex4}
\usepackage{amsmath, graphicx}
\usepackage{color}

\newcommand{\Cp}{${\bf C_p\,}$}
\newcommand{\D}{${\bf D\,}$}
\newcommand{\K}{${\bf K\,}$}
\newcommand{\LL}{${\cal  L\,}$}

\newcommand{\be}{\begin{equation}}
\newcommand{\ee}{\end{equation}}
\newcommand{\ba}{\begin{eqnarray}}
\newcommand{\ea}{\end{eqnarray}}

\begin{document}

\title{Dynamical fluctuations close to Jamming versus Vibrational Modes : a
pedagogical discussion illustrated on hard spheres simulations, colloidal and
granular experiments}

\author{Silke Henkes}
\affiliation{Physics Department, Syracuse University, Syracuse, NY 13244, USA}

\author{Carolina Brito}
\affiliation{Inst. de Fisica, Universidade Federal do Rio Grande do Sul, CP 15051, 91501-970, Porto Alegre RS, Brazil}

\author{Olivier Dauchot}
\affiliation{EC2M-Gulliver, ESPCI-ParisTech and CNRS UMR 7083, 75005 Paris, France}

%*******************************************************************************
\begin{abstract}
Studying the jamming transition of granular and colloidal systems, has lead to a
proliferation of theoretical and numerical results formulated in the language of
the eigenspectrum of the dynamical matrix for these disordered system. 
Only recently however, these modes have been accessed experimentally in
colloidal and granular media, computing the eigenmodes of the covariance matrix
of the particle positions. At the same time new conceptual and methodological
questions have appeared, regarding the interpretation of these results.
In the present paper, we first give an overview of the theoretical framework
which is appropriate to discuss the interpretation of the eigenmodes and
eigenvalues of the correlation matrix in terms of the vibrational properties of
these systems.
 We then illustrate several aspects of the statistical and data analysis
techniques which are necessary to extract reliable results from experimental
data. Concentrating on the case of hard sphere simulations, colloidal and
granular experiments, we discuss how to test the existence of a metastable
state, the statistical independence of the sampling, the effect of the
experimental resolution, and the harmonic hypothesis underlying the approach, 
highlighting both its promises and limitations. 
\end{abstract}
\maketitle

\tableofcontents

%*******************************************************************************
\section{Introduction}
\label{intro}
%*******************************************************************************
Amorphous systems such as structural glasses, colloids, emulsions or granular
matter still lack a satisfying description.
This is particularly apparent when one considers the low temperature properties
of glasses~\cite{Phillips1978, Nakayama-RepProgPhys2002} or the sometime
intriguing rheological properties of athermal
amorphous systems, such as foams\cite{Katgert-EPL2010, Tighe-PRL2011} or
granular matter~\cite{vanKecke-review2010, Dijksman-PRL2011}.
Of particular importance is the understanding of what guarantees the mechanical
stability of such systems \cite{Alexander1998, wyart_thesis} . 
 The informations about the rigidity of a solid against collective particle motions are 
contained in the density of the vibrational modes $D(\omega)$: a system is stable -- 
at least linearly -- if there are no unstable modes. 
In a continuous isotropic elastic medium, the invariance by translation implies
that the vibrational modes are plane waves and the density of vibrational modes
$D(\omega)$ follows the Debye law $D(\omega) \sim
\omega^{d-1}$\cite{Ashcroft-Mermin}, where $d$ is the space dimension.
By contrast, disordered solids exhibit common low-frequency vibrational
properties that are completely unlike those of crystals. Disordered atomic or
molecular solids generically exhibit a \lq\lq boson peak\rq\rq,
where many more modes appear than expected for
sound~\cite{Chumakov-PRL200, Wyart_EPL2005, Xu-PRL2007, PRL_NIPA_2010}. 
Several empirical facts suggest that the presence of these excess modes
is related to many of the original properties of amorphous
solids~\cite{Baldi-PRL2009, Tanaka-NatureMat2008, Nakayama-RepProgPhys2002}.

The jamming transition of frictionless granular materials lies at the threshold
of mechanical stability, also known as the isostatic point. As a result of this
equivalence, an excess of low-frequency modes vibrational arises and gives rise to a plateau
in the density of states at low frequencies\cite{Silbert05}. This suggests that
the zero-temperature jamming transition may provide a framework for
understanding at least some of the properties of amorphous solids. As a result of this
an extensive theoretical and experimental effort has been conducted towards the
study of the vibrational properties of various amorphous systems close to
jamming. 

For frictionless repulsive soft spheres, the vibrational modes are obtained by
diagonalizing the Hessian $K_{ij}=\frac{\partial^{2} V}{\partial r_{i} \partial r_{j}}$,
defined as the matrix of the second derivatives of the  pair interaction
potential $V(\{r_{i}\})$ with respect to the particles displacements around a
given metastable state $\{r_{i}^{0}\}$ obtained numerically through conjugate
gradient minimization.  Above the jamming transition, it was shown that the
onset frequency $\omega^{*}$ of the anomalous modes scales linearly with
the distance from isostaticity, and that it is related to the existence of a
diverging length scale in the response  to an
external perturbation\cite{Wyart05,Ellenbroek06}.
Beyond the frictionless, non-dissipative case, the density of states has also
been studied for non-spherical\cite{Mailman09, Zeravcic09, Schreck10} and for
frictional\cite{Somfai07,Henkes10} soft particles. In the specific case of hard
spheres below jamming, for which the potential is singular, an effective potential is first
derived from the transfer of momentum during collisions in order to apply the
same procedure~\cite{BritoWyart_EPL2006}.

Experimentally, the density of states is traditionally accessed through light
scattering techniques~\cite{Hansen, Nakayama-RepProgPhys2002,  Greaves-AP2007} 
or by computing the Fourier transform
of the velocity autocorrelation function~\cite{Keyes97}, as was recently done in
thermal soft sphere simulations~\cite{Xu09} and in some granular
experiments~\cite{Daniels_MM11}. Such methods do not give access to the spatial
structure of the vibrational modes and only apply for thermally equilibrated
systems. This is not the case of many systems of interest such as  granular
systems, foams and suspensions of large colloidal particles. 
For such systems, recent advances in experimental techniques now allow tracking
the real space dynamics of the individual  particles\cite{Weeks2000}. In principal one
should thus be able to follow the same procedure as in simulations, namely
average the particle positions in order to obtain a reference state and then
diagonalizing the Hessian around this state. However, the interaction potential
$V(\{r_{i}\})$ is usually unknown. For granular systems or foams, the elastic
part of the interaction is well understood but the pairwise dissipation, either
static friction or viscous damping, does not derive from a potential. In
colloidal suspensions, the presence of hydrodynamics interaction prevent from
deriving a simple two-body interaction rule. Also the experimental resolution 
may be a limiting factor to perform such an analysis.

An alternative approach is to follow the dynamics and perform a Principal
Component Analysis (PCA) of
the covariance matrix of the positions around the reference state of interest.
This technique allows to extract the dominant components of the fluctuations --
in the sense that they concentrate most of the correlated motion. It has been
applied successfully in a range of fields, from the pathways of protein folding
\cite{Micheletti-PCA2004, Petitjean20101790} to financial mathematics
\cite{Laloux-PRL1999}. Whatever the underlying dynamics are, the PCA acts as
a filter which separates the correlated significant part of the fluctuations from
the uncorrelated noise.

For systems at thermal equilibrium, we shall recall that the eigenmodes of the
covariance matrix of the positions are identical to those of the Hessian, and
that the associated eigenvalues are related through a simple relation. Following
this line several teams have recently attempted to extract the density of state
from various experimental systems such as  hard-sphere
colloids \cite{PRL_Colloids2010,Kaya10,Yodh11a,Yodh11b} and vibrated granular
materials \cite{Brito_SM2010}. While contributing to this
effort~\cite{BritoWyart_EPL2006, PRL_Colloids2010, Brito_SM2010}, 
the authors of the present paper have realized that both the interpretation of
the PCA analysis and its practical implementation requires some attention. 

In this paper, we would like to discuss both the promise and the limitations of this
approach, as well as the set of hypotheses it depends on and the interpretation
one can give to the experimental results. We pay particular attention
to the influence of the real dynamics on the interpretation of the modes, and
also point out the fundamental differences between equilibrium and
non-equilibrium systems. We discuss the differences between the idealized spring
network to which the eigenfrequencies refer to  and the actual physical system
by developing the concept of \emph{shadow system}, first introduced
in~\cite{PRL_Colloids2010}. We illustrate the discussion with simulations of
jammed soft spheres obeying an over-damped Langevin dynamics.
On the way, we will see that 
\begin{itemize}
\item (i) whenever the modes can be computed, they indeed closely match those
of the Hessian and are meaningful to the description of the dynamics, 
\item (ii) the eigenspectrum of the covariance matrix of the position allows to
compute the density of state only in the long time limit and if equipartition
holds,
\item (iii) beyond the density of states,  the covariance matrix of the positions is a
rich source of information regarding the local rigidity properties of the system. 
\end{itemize}

We would also like to provide some practical guidelines, which we hope will
spare some time to those who would like to follow the above approach. We again
 illustrate our purpose with several systems of interest. First, we use hard sphere
simulations to address statistical issues, while experiments with
granular systems close to jamming illustrate the possible anharmonicity
of the dynamics. Finally, two experiments with colloidal suspensions~\cite{PRL_NIPA_2010,PRL_Colloids2010} illustrate
convergence and resolution issues, in particular the problems that arise from an insufficient number of samplings~\cite{PRL_Colloids2010}.

At this point let us state that the present paper is written in a pedagogical
spirit. It does not so much contain revolutionary ideas as instead the first
reasonably complete and coherent treatment of the conceptual and practical
issues surrounding this approach. We hope to have injected some wisdom on the
use of PCA applied as to soft matter physics, and also of the broad relevance of
the Mar\u{c}enko-Pastur theorem. It is the result of many discussions
with those colleagues of ours who pioneered this kind of delicate analysis,
and we are grateful to them. It is our best wish that this paper provide an entry
point to those entering the game.

%*******************************************************************************
\section{Theoretical background}
\label{theory}
%*******************************************************************************
In this section we present rigorous derivations of the link between the
covariance  matrix of the positions and the Hessian matrix. Let's consider a
system of $N$ particles following trajectories $|r(t)\rangle$ and let's assume
one can define a statistical reference state $|r^{0}\rangle = 
\left<|r(t)\rangle \right> $ during a time interval $T$, where   $\left<
~.~\right>$  denotes the average over time $T$. One then introduces the
displacement vector $|\delta r(t) \rangle=  | r(t) \rangle-  | r^{0} \rangle$,
the Hessian matrix $K_{ij}=\frac{\partial^{2} V}{\partial r_i  \partial r_j}$,
defined as the matrix of the second derivatives of the  pair interaction
potential $V(|r\rangle)$ with respect to the particles displacements around the
reference state $|r^{0}\rangle$, and the covariance matrix of these
displacements:

\begin{equation}
{\bf C_p} = \left< ~ |\delta  {r(t)} \rangle ~\langle \delta {r(t)}|~ \right>.
\label{Cp}
\end{equation} 

\noindent 
Although the displacement vector $|\delta r(t) \rangle$ depends on $t$, we will
hereafter omit this dependence to simplify the notation. \Cp is the empirical
correlation matrix (i.e. on a given realization of the trajectories), that one
must not confuse with the true correlation matrix of the underlying statistical
process. While in this first part we will assume that $T$ is much larger than $N$
ensuring the equivalence between time and ensemble averages and thereby the
equality between the empirical and the true correlations, we shall see
later that this is usually not the case in practical situations. Note that ensemble average
here refers to the statistical average over the thermal fluctuations around a given glassy state.
We will then have to refer to the Mar\u{c}enko-Pastur theorem~\cite{MarcenkoPastur}
and its generalizations to discuss the convergence properties of the spectrum of \Cp
to the one of the true correlation matrix.

\Cp is a real symmetric matrix, that can always be diagonalized. We
will see now how to relate its modes and eigenvalues to those of the Hessian,
depending on the underlying microscopic dynamics. We first discuss two important
classes of equilibrium dynamics, namely inertial Newtonian and fully over-damped
Langevin dynamics. The generalization to the case of an inertial and partially
damped dynamics in the presence of a colored noise will allow us to discuss the
case of athermal and mechanically excited systems. 
 
\subsection{Equilibrium systems}
%*******************************************************************************
\label{equilibrium}
For systems in thermal equilibrium, statistical physics tells us that one can forget the
dynamics and replace it by Gibbs statistics. One defines the partition function 
$Z \propto \int  \exp(- \beta V) d |\delta r \rangle$ and in the harmonic
approximation $V= \langle \delta r | {\bf K} |\delta r \rangle / 2$, so
that it is straightforward to compute the correlation functions from the
partition function. In particular, the correlation of the displacements reads
\begin{equation}
{\bf C_p} = k_BT~ {\bf K}^{-1}.
\label{C_vs_K}
\end{equation}
\noindent 
Hence for equilibrium dynamics, the eigenvectors of \Cp are simultaneously
eigenvectors of \K, while their eigenvalues are simply inversely proportional.
We shall now illustrate how one can recover the above relationship in the cases
of Newtonian dynamics and over-damped  Langevin dynamics. The goal of this
calculation is twofold. First it will point at the underlying hypothesis
satisfied by equilibrium dynamics. Second it will provide a simple sketch of the
more intricate calculation, which we will have to perform to discuss the case of
non equilibrium mechanically excited systems.

\subsubsection{Newtonian dynamics}
%***************************************
\label{theoretical_newtonian}
\emph{In the harmonic approximation}, that is linearizing the dynamics around the
reference state $|r^0\rangle$, one has for non dissipative dynamics
\begin{equation}
 |\delta  \ddot{r} \rangle + \frac{\bf K}{m} |\delta  {r} \rangle = 0.
\label{newtonEq}
\end{equation}  
\noindent
This is the first and crucial approximation underlying the whole approach.
Since the reference state is supposed to be mechanically stable, the eigenvalues
$\kappa_q$ of \K are positive. Introducing 
${\bf D}={\bf K}/m$, also called the dynamical matrix, and where for simplicity we
have supposed that all particles have a same mass $m$.  The eigenmodes of ${\bf
D}$, also called the vibrational modes of the system, are then defined by  ${\bf
D} |\lambda_q\rangle = \omega_q^2 |\lambda_q\rangle$, where 
$\omega_q=\sqrt{\kappa_q/m}$ are the vibrational frequencies.\\

\noindent
The solution of Eq.(\ref{newtonEq}) is given by:
\begin{equation}
|\delta  {r} \rangle = e^{-i\sqrt{\bf D}t}  |\delta  {r(0)} \rangle,
\end{equation}
\noindent
where  $|\delta  {r(0)} \rangle$ is the displacement field at time zero.
Replacing this solution in Eq.(\ref{Cp}) and writing  $|\delta {r(0)} \rangle$
in the eigenbasis of \D, $\{|\lambda_q\rangle\}$, one easily shows (see appendix
\ref{appendix_Newton} for derivation)  that \emph{in the long time limit} (long
compared to the inverse of the minimal gap between adjacent frequencies), \Cp
is diagonal in the eigenbasis of \D:   
\begin{equation}
{\bf C_p}|\lambda_q \rangle =\alpha_q^2  ~ |\lambda_q\rangle,
\label{eigen_Eq_Newton}
\end{equation} 
\noindent 
where $\alpha_q=\langle \lambda_q|\delta {r(0)} \rangle$ is the amplitude of the
initial condition on the mode $|\lambda_q\rangle$.\\

\noindent
Now comes the third key ingredient: at equilibrium, the initial condition
is thermalized and \emph{the energy is equally distributed among the modes}. Each mode
of frequency $\omega_q$ and amplitude $\alpha_q$ carries an energy $m\alpha_q^2
\omega_q^2/2=k_BT/2$.  Hence the eigenvalues $\lambda_q$ of \Cp and the
vibrational frequencies $\omega_q$ of the dynamical matrix are related through
\begin{equation}
 \lambda_q = \alpha_q^2 = \frac{k_BT}{m\omega_q^2}; \quad \omega_q = \sqrt{
\frac{k_BT}{m\lambda_q}}
\label{evalues_Newton1}
\end{equation}
Practically, we have seen that for a Newtonian dynamics at equilibrium, the
diagonalization of \Cp does provide the vibrational modes of the system, the
frequencies of which are proportional to the square root of the inverse of the
eigenvalues of \Cp. The largest eigenvalues of \Cp, that is the most coherent
motion corresponds to the softest modes of the system.

\subsubsection{Overdamped Langevin dynamics}
%****************************************************
Again, \emph{in the harmonic approximation}, one obtains the following
Langevin equation for an over-damped dynamics:
\begin{equation}
|\delta \dot{r}(t)\rangle =  - \frac{{\bf K}}{\mu} |\delta{r}\rangle +
\frac{1}{\mu}|\eta(t)\rangle, 
\label{LangevinEq}
\end{equation}
\noindent
where $\mu$ is the viscous damping and $|\eta(t)\rangle$ is a white noise of
amplitude $\Gamma$: $\langle \eta(t')  |\eta(t'')\rangle = \Gamma \delta(t'-t'')
$, with $\Gamma=2\mu k_BT$ since we consider equilibrium dynamics, for which
the \emph{fluctuation-dissipation relation} holds.\\
 
\noindent
To discuss this case, we introduce the operator  ${\cal  L} = {\bf K}/\mu$ whose
eigenvalue equation is
 ${\cal L}|\lambda_q\rangle = \kappa_q /\mu |\lambda_q\rangle$. The
eigenvalues have the dimension of an inverse time, which one interprets as
the relaxation time $\tau_q =\mu/\kappa_q=\mu/(m\omega_q^2)$ of the system
along the eigenmode $|\lambda_q\rangle$.\\

\noindent
The solution of eq.(\ref{LangevinEq}) can be written as
\begin{equation}
 |\delta{r}\rangle = e^{-{\cal L} t} |\delta  {r(0)} \rangle + 
\frac{1}{\mu}\int_0^{t}e^{-{\cal L} (t-t')}|\eta(t')\rangle dt' 
\end{equation}
\noindent
Following the same path as for the Newtonian dynamics, we show in the appendix 
\ref{appendix_Langevin} that \Cp is diagonal in the eigenbasis of \LL:
\begin{eqnarray}
C_p |\lambda_q \rangle =  \left[ \left( \alpha_q^2 - \frac{\Gamma \tau_q}{2\mu^
2}\right)  e^{-2t/\tau_q} + \frac{\Gamma \tau_q }{2\mu^ 2}\right] |\lambda_q
\rangle,
\label{op_Cp}
\end{eqnarray}
\noindent
where $\alpha_q=\langle \lambda_q|\delta {r(0)} \rangle$ is the amplitude of the
initial condition on the mode $|\lambda_q\rangle$. To derive the above relation,
it is assumed that the initial conditions as well as the noise components on the
modes are both uncorrelated. Note that under such assumptions, \Cp is diagonal
in the basis of $\cal L$, even for finite time. However it is only \emph{in the long
time limit} $t>>\tau_q$ for the largest $\tau_q$ that one recovers the simpler
expression
\begin{eqnarray}
\lambda_q  = \frac{\Gamma \tau_q}{2 \mu^2} = \frac{k_BT}{m\omega_q^2},
\label{evalues_Langevin1}
\end{eqnarray}
\noindent
where we have used the fluctuation-dissipation theorem $\Gamma=2 \mu k_B T$.\\

Practically, we have seen that for an overdamped Langevin dynamics, the
diagonalization of \Cp again allows us to compute the vibrational modes of the
system. The eigenvalues $\lambda_q$ are related to the relaxation time $\tau_q$
of the dynamics in each mode, which are themselves related to the oscillating
frequencies $\omega_q$ of the undamped system.

\subsection{Athermal systems}
%*********************************************************************
\label{non-equilibrium}
For many systems of interest, such as grains, foams, suspensions of large
particles, the thermal fluctuations are orders of magnitude too weak to drive
the dynamics. Such systems are thus generically out of equilibrium, the
interactions lead to dissipation and some external forcing must be provided to
drive the dynamics.  The central thermal equilibrium relation
equation~(\ref{C_vs_K}) is based on assuming both ergodicity of the fluctuations
around the reference state during the observation period, and an underlying
thermal canonical ensemble. Both assumptions are a priori violated in a
non-equilibrium system. Also we have seen above that several hypothesis are
necessary to derive a relation between the eigenvalues of \Cp and those of the
Hessian. Most of these hypotheses are related to the correlation properties of
the noise. These could be extended to non equilibrium situations. However, one
key assumption is the equipartition of the energy on the modes. This property is
very specific to equilibrium. It is well known that in out of equilibrium
situations driven by a macroscopic forcing, the energy is in general not
equi-distributed. On the contrary, the generic situation is that non linearities
drive a cascade of energy from the large scale to the dissipative scale.
For most systems, the spectrum of the energy fluctuations is not even known
theoretically.

Here we conduct the same kind of analysis as above for a homogeneously driven
dissipative system. Vibrated grains experiments are typical realizations of this
situation.  To generalize from the two equilibrium cases of Newtonian dynamics and
Langevin dynamics, we now consider a non-equilibrium system with both inertia
and damping and a colored noise spectrum. We choose to stick to a Langevin type
description of the dynamics, where  the damping is linear and single particle,
as appropriate for particles in a newtonian fluid bath, but not necessarily for
a large scale mechanical excitation. The generic equations of motion for the
linearized dynamics around a reference state  $|r^0\rangle$ is then:
\begin{equation}
 m|\delta \ddot{r}(t) \rangle + \mu |\delta \dot{r}(t) \rangle = -{\bf K} |
\delta r(t) \rangle + |\eta(t)\rangle
\label{eq:stochmech}
\end{equation}
\noindent
where $\mu$ is the viscous damping and $|\eta(t)\rangle$ is a \emph{colored}
noise defined in the basis of the modes as
\begin{align}
& |\eta(t)\rangle=\sum_{q} \eta_{q}(t) |\lambda_{q}\rangle \quad \nonumber \\
& \text{with} \quad \langle \eta_{q}(t) \eta_{k} (t') \rangle = \Gamma_{q}
\delta_{q k} \delta(t-t'). 
\end{align}
We assume that the noise correlations are fully described by their first and second
moments, i.e. a normal distribution.
\noindent
Here $\Gamma_{q}$ is the amount of energy which flows into the system through
non-equilibrium processes along this mode. It should be emphasized that in
general this function is unknown and difficult to determine. \\

Solving the dynamics projected on the modes of \K, using the same notation as
for the Langevin case, ${\bf K} |\lambda_{q} \rangle = \kappa_{q} |\lambda_{q}
\rangle$  (see appendix~\ref{appendix_stochmech}), we find that \Cp is diagonal
in the eigenbasis of \K. 
At finite time one obtains complicated eigenvalues where the damping and the
oscillatory component of the dynamics interplay and one must distinguish the low
($(\mu/m)^2 < 4 \kappa_q /m$) versus the strong ($(\mu/m)^2 > 4 \kappa_q /m$)
damping limits (equations~\ref{eq:ev1} and~\ref{eq:ev2} of the appendix). 
We have also assumed that noise and
initial conditions do not cross-correlate, and that in ensemble-average, the
initial conditions of different modes are independent of each other.  These
conditions are in principle not so easy to satisfy if the noise is produced by a
regular external excitation with a well-defined coupling to the modes.
Fortunately, \emph{in the long time limit} the system loses memory of its initial
conditions and one recovers  an expression similar to that of the equilibrium
result 
\begin{equation}
 C_{p}|\lambda_{q} \rangle = \frac{\Gamma_{q}}{2 \mu \kappa_q} |\lambda_{q}
\rangle 
\end{equation}
However one does not get rid of the violation of equipartition: the eigenvalues
of \Cp are the ratio of two amplitudes, one being the amount of energy the
external forcing has put on the mode, the other being proportional to the mode
stiffness. Hence the eigenvalues of \Cp do \emph{not} give access to the density of
states of the system.
 
%*******************************************************************************
\section{Interpretation}
\label{interpretation}
%*******************************************************************************
We have seen in the above section that for systems in thermal equilibrium, one
can in principle safely use the covariance matrix \Cp of the positions around an
equilibrium state in order to obtain the  stiffness matrix \K, or equivalently the
dynamical matrix \D, and the eigenfrequencies of the system using the relations:
\begin{eqnarray}
{\bf K}&=&k_B T\, {\bf Cp}^{-1}
\label{KCp}
\end{eqnarray}
\begin{eqnarray}
\omega_q^2&=&\frac{k_BT}{m\lambda_q}.
\label{wqlq}
\end{eqnarray}
\noindent
However, real systems can be vastly more complicated than the idealized
Newtonian or Langevin equilibrium systems described above. 
Even a model system of jammed harmonic frictionless soft spheres obeying either
Newtonian dynamics or treated with a conjugate gradient algorithm exhibits strong
nonlinearities when approaching the jamming transition~\cite{vanHecke_MM11, O-HernPRL}.
The description of purely repulsive
soft or hard spheres below jamming requires the introduction of an effective
potential as defined by the transfer of momentum among the
particles~\cite{BritoWyart_EPL2006}.  In colloidal systems, electrostatic
charges and the effect of solvent can significantly alter the pair interaction
potential including the development of long range interactions. Finally, we have
seen that for out of equilibrium systems the possible violation of the
equipartition relation prevents us from using relations~(\ref{KCp})
and~(\ref{wqlq}).  

It is thus of primary importance to clarify the physical interpretation one
can give to the eigenmodes and eigenvalues of \Cp. As we shall see below, there
are two strategies. One may introduces a ``shadow system'', which is
\emph{by definition} the thermally equilibrated system of which 
$k_B T\, {\bf Cp}^{-1}$ {\it is} the rigidity matrix. As long as the hypothesis
underlying relations~(\ref{KCp}) and~(\ref{wqlq}) remains reasonable for the
system of interest, this shadow system should in principle have the same
eigenspectrum as the real one.

If these hypotheses are not valid, in particular if the harmonic approximation
or the equipartition are strongly violated, one should alternatively stick
to Principal Component Analysis without expecting to relate the eigenvalues
of \Cp to the vibrational properties. Note that the comparison with thermal
systems of reference can still be done, but computing the spectrum of \Cp
from the density of states of the thermal system instead of trying to compute the
density of state of the athermal system from \Cp. It can also be of interest, as we
will illustrate below,  to define, on the basis of the experimental data, a system of
reference, the correlations of which have been suppressed.

\subsection{The \lq\lq Shadow system\rq\rq : vibrational modes and density of
states}
%*******************************************************************************
\label{shadow}

\begin{figure*}[t!]
\centering
\includegraphics[width = 0.60\columnwidth,trim = 30mm 10mm 25mm 5mm, clip]{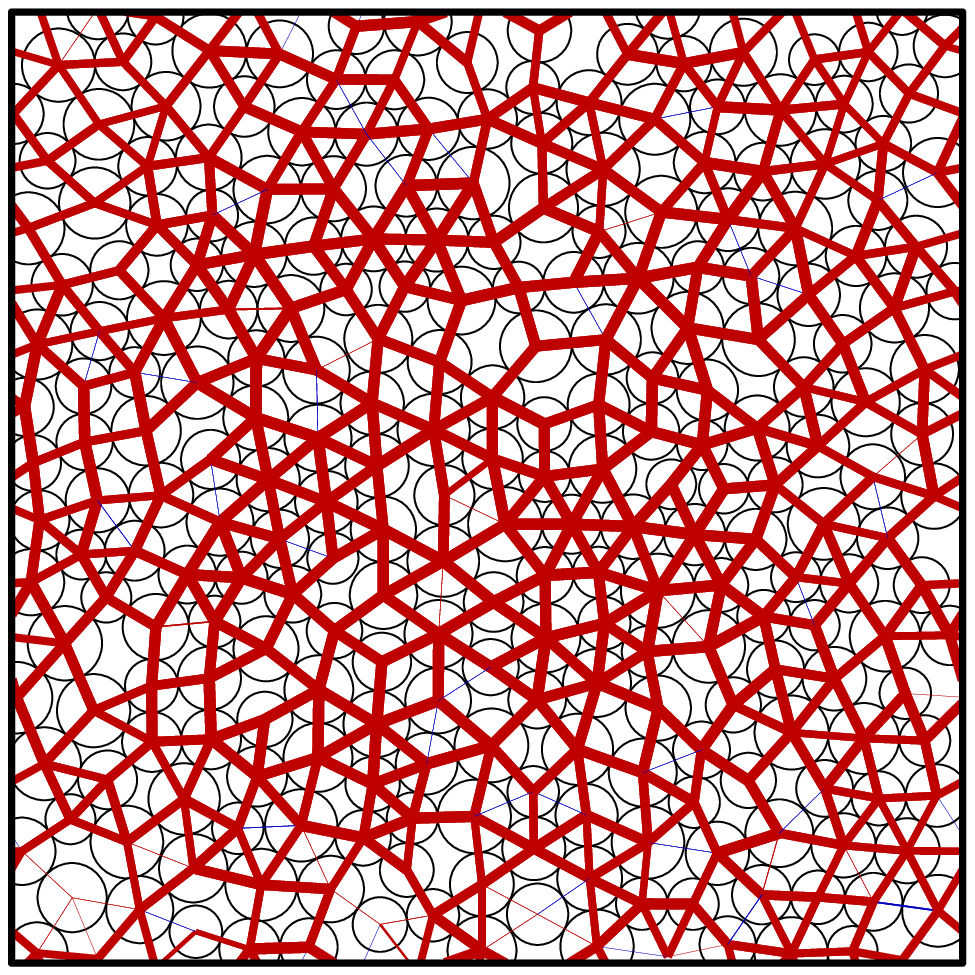}
\includegraphics[width = 0.60\columnwidth,trim = 0mm 0mm 12mm 5mm,clip]{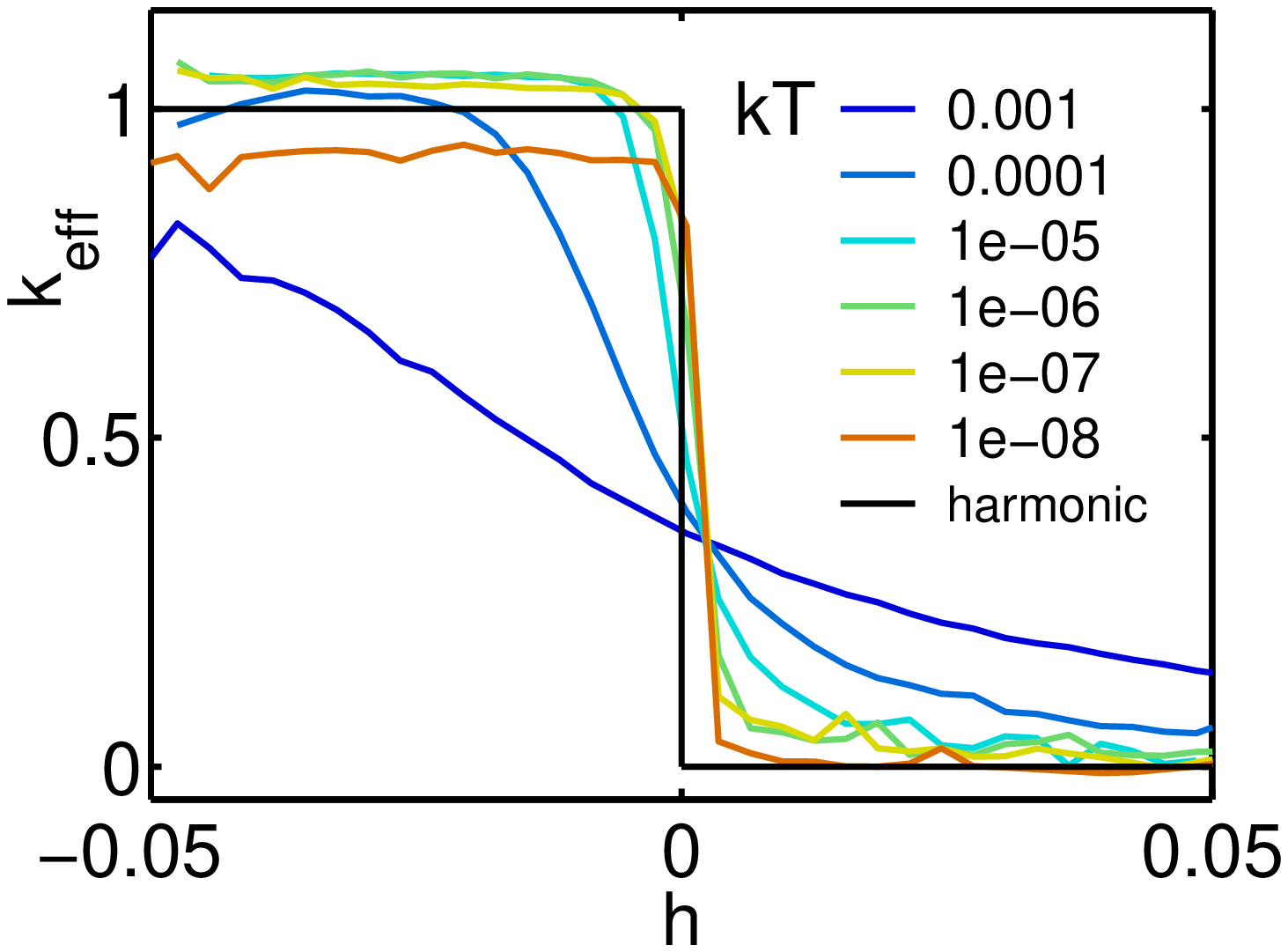}
\includegraphics[width = 0.60\columnwidth,trim = 0mm 0mm 12mm 5mm,clip]{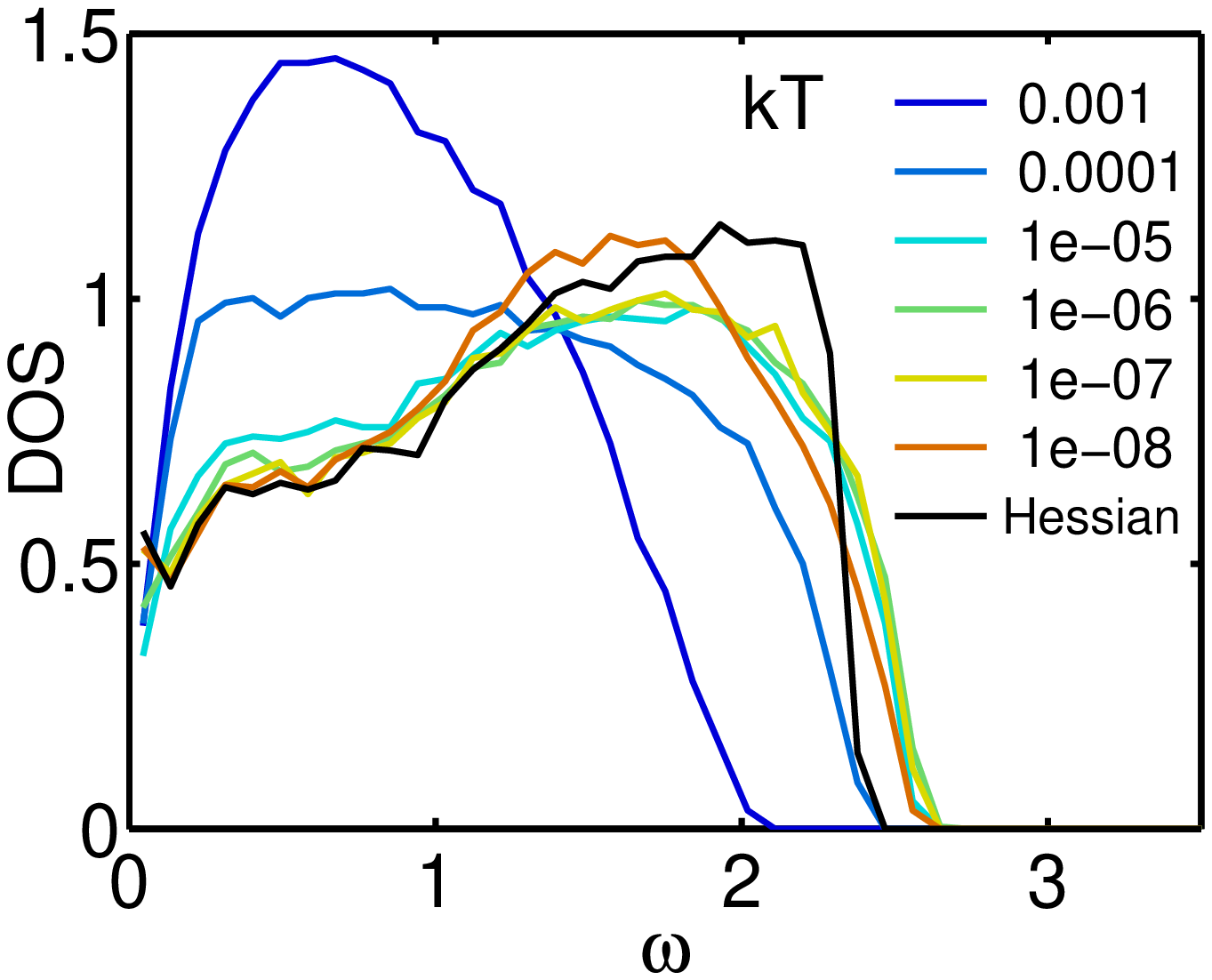}
\caption{(color online) Left: Shadow system for a brownian system composed of soft harmonic spheres slightly
above jamming at $\phi=0.86$ and $kT=10^{-8}$. The contacts are plotted with widths proportional to their
effective stiffness $k_{\text{eff}}$, red with positive $k_{\text{eff}}$ and
blue with negative $k_{\text{eff}}$. Center: Dependence of the 
effective stiffnesses $k_{\text{eff}}$ on the interparticle gap for different temperatures $kT$ and $\phi=0.86$; compare
to the step function for the harmonic potential set in the simulation. Right: Density of
states $D(\omega)$ computed from the eigenspectrum of \Cp for the same range of temperatures, compared to the one
derived from the Hessian of the soft sphere potential.}
\label{fig:shadow_system_SS}
\end{figure*}

Let us define the shadow system explicitly. Suppose some experimental system
of colloidal particles moving around a metastable configuration ${|r^0 \rangle}$.  
Once computed the effective spring constants $k_{ij}^{\text{eff}}$ between
particles from the empirical inverse stiffness matrix $\K^{\text{eff}}=kT {\bf
C_p}^{-1}$, the shadow system is defined as the Hamiltonian system of harmonic
springs with stiffnesses $k_{ij}^{\text{eff}}$, strung between particles of mass $m$
sitting in the metastable configuration ${|r^0 \rangle}$. 
Note that the  \emph{shadow system} must not be confused with the \emph{model
system} the physicist has in mind when discussing his observations on the
\emph {experimental system}. In the present case, despite possible complicated
electrostatic and hydrodynamic interactions, the colloidal system is considered
as a good experimental realization of over-damped soft spheres in a thermal bath
- the model system the physicist has in mind. Whereas the model system obeys 
an overdamped dynamics, the shadow system obeys the Hamiltonian dynamics of 
perfectly harmonic springs.

For thermal systems at equilibrium, the spectrum of \Cp can be interpreted as
the vibrational modes of the shadow system and both the shadow system and the
model system share the same density of states. However one must remember that the
\emph{vibrational properties} of the model system are still different from those
of the shadow system. First, even in the linear regime, it is well known that
resonance frequencies of a system of springs are shifted by damping and
eventually completely eliminated in the over-damped limit. Second, the model
system may well escape the linear regime. This is what happens for frictionless
disks close to jamming, because of the vanishing range of linear response, as
was shown recently~\cite{O-HernPRL}. In the following we illustrate this last
remark with two computational ``experimental systems''. The first system
consists of poly-disperse packings of thermal frictionless soft spheres
interacting through a harmonic potential, and simulated with over-damped
Langevin dynamics~\cite{Tildesley} close to the jamming transition. The second
one consists of thermal hard sphere packings simulated with event-driven
Newtonian molecular dynamics just below the jamming transition. Initial states
are obtained by shrinking simulated packings at jamming by a factor $\delta \in
[10^{-6} - 10^{-2}]$  which controls the distance from
jamming~\cite{BritoWyart_EPL2006}.

Figure~\ref{fig:shadow_system_SS} displays the effective stiffnesses obtained for the
soft sphere simulations slightly above jamming, at a very low temperature. We find
effective stiffnesses consistent with the original connectivity and the original
stiffness, which is a constant, here set to $1$ for the harmonic potential. The only exceptions
are occasional weak interactions with cage neighbors which are not contacts.
As can be seen in Figure~\ref{fig:shadow_system_SS} center, at low temperatures, 
the stiffness as a function of overlap remains close to the $T=0$ step function.
At low temperatures, as one would have expected for soft spheres above jamming, the
shadow system is reasonably close to the ``experimental'' one. As a
result the density of states $\D(\omega)$ computed from the Hessian matrix and
the one computed from the spectrum of \Cp closely match (see figure~\ref{fig:shadow_system_SS}, right).

When the temperature is increased, still remaining deeply within the glassy regime, we
find that the stiffness as a function of the overlap broadens away from the $T=0$ step function,
see Figure~\ref{fig:shadow_system_SS} bottom left. Equally, we find that the density
of states obtained from \Cp begins to differ substantially from the Hessian density of states.
The larger temperatures allow the system to explore the potential beyond the range
of linear response. 

\begin{figure}
\centering
\hspace{0.015\columnwidth}
\includegraphics[width = 0.42\columnwidth,trim = 20mm 10mm 25mm 5mm,clip]{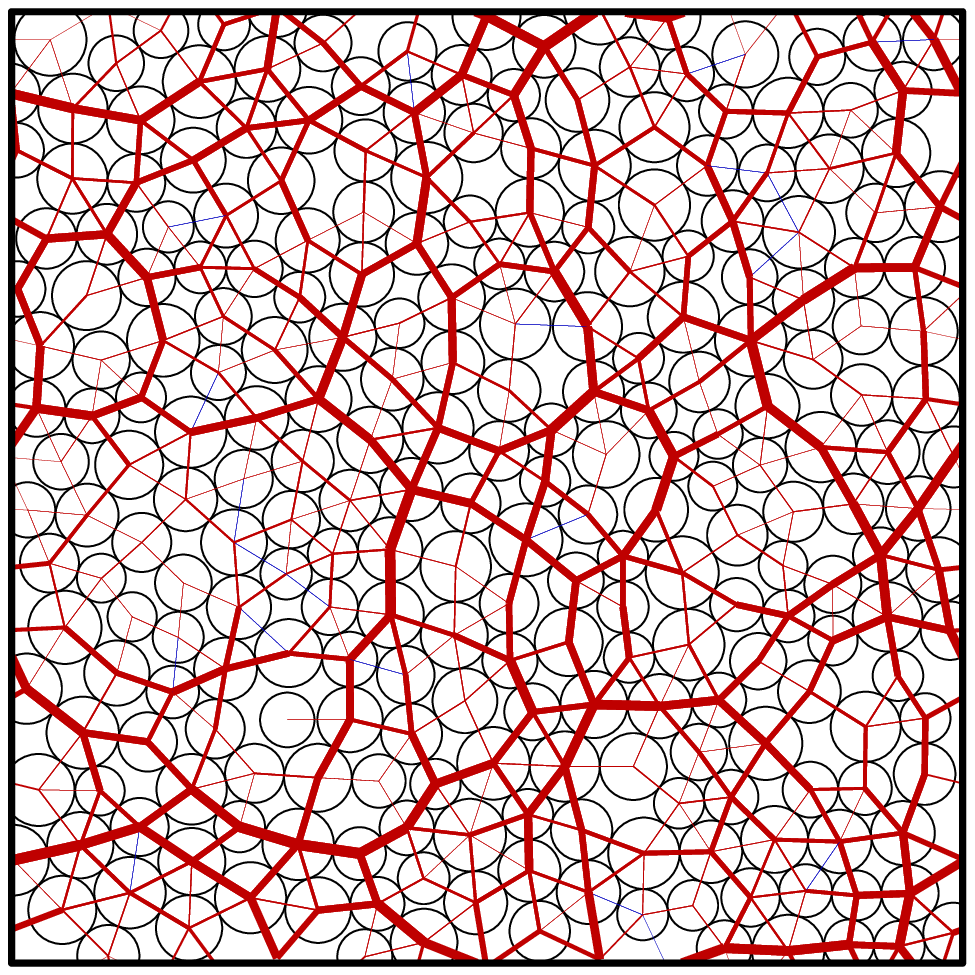} \hspace{0.01\columnwidth}
\includegraphics[width = 0.52\columnwidth,trim = 0mm 0mm 5mm 0mm,clip]{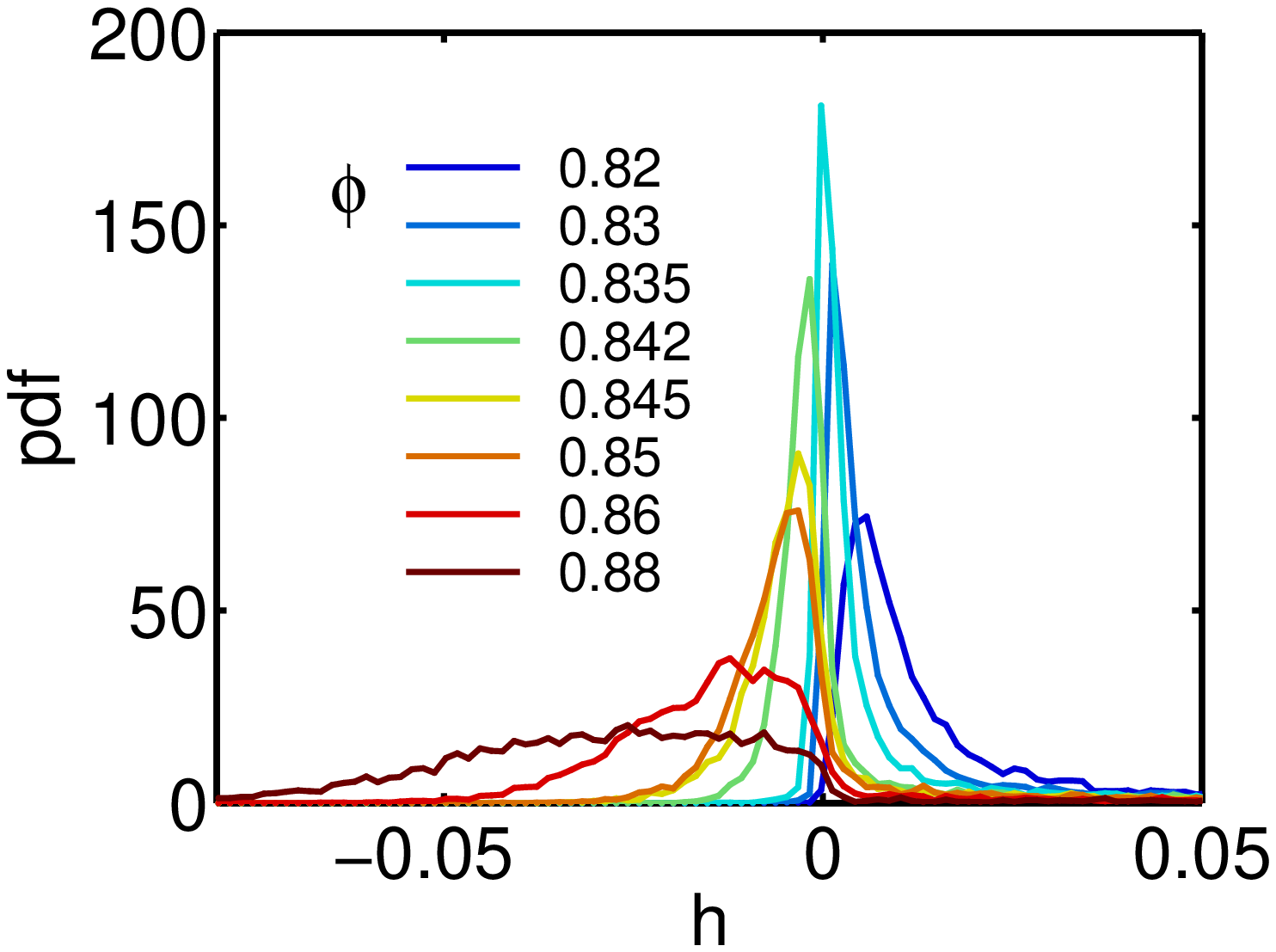}
\includegraphics[width = 0.49\columnwidth,trim = 5mm 0mm 7mm 5mm,clip]{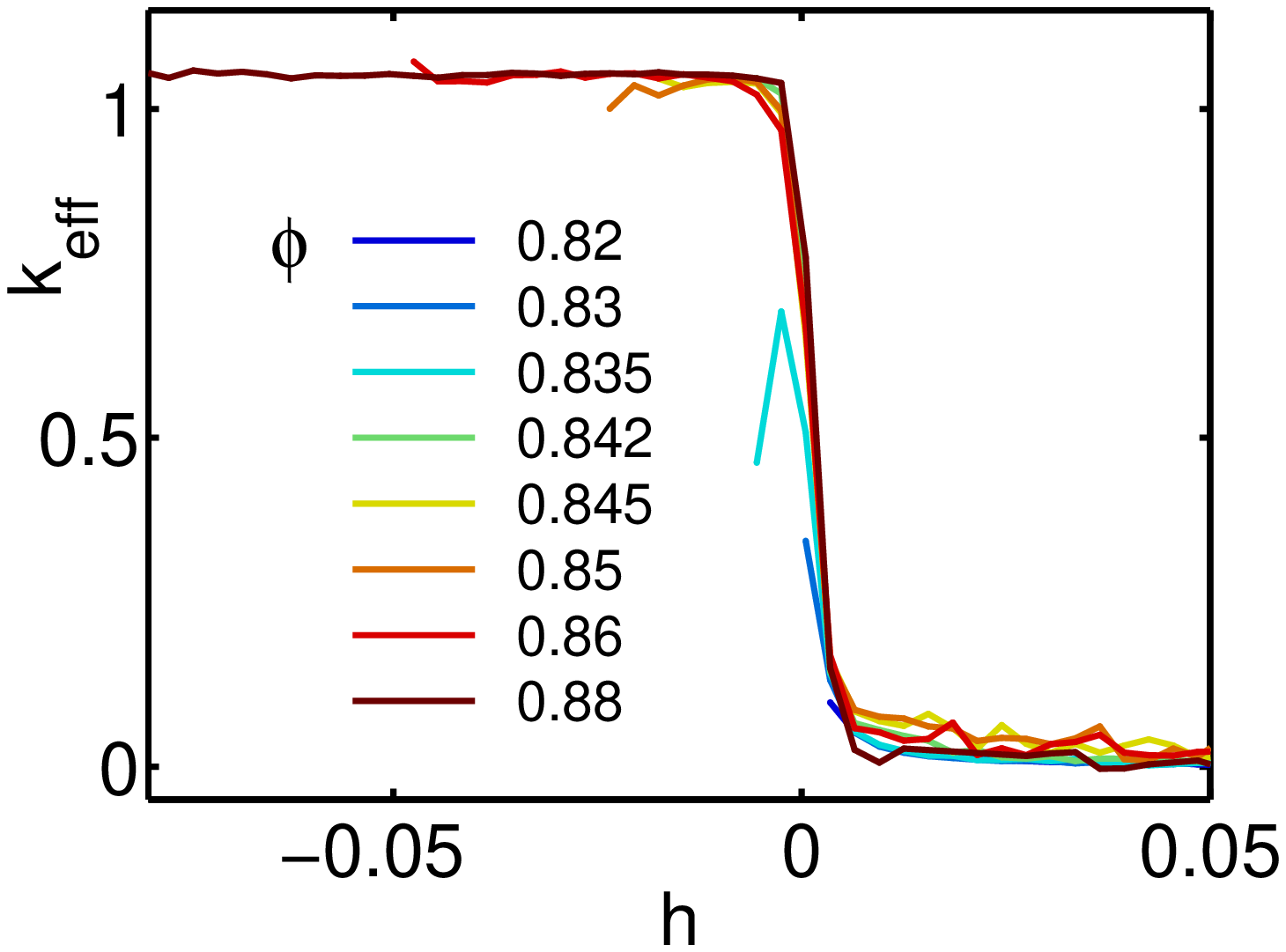}
\includegraphics[width = 0.49\columnwidth,trim = 0mm 0mm 12mm 5mm,clip]{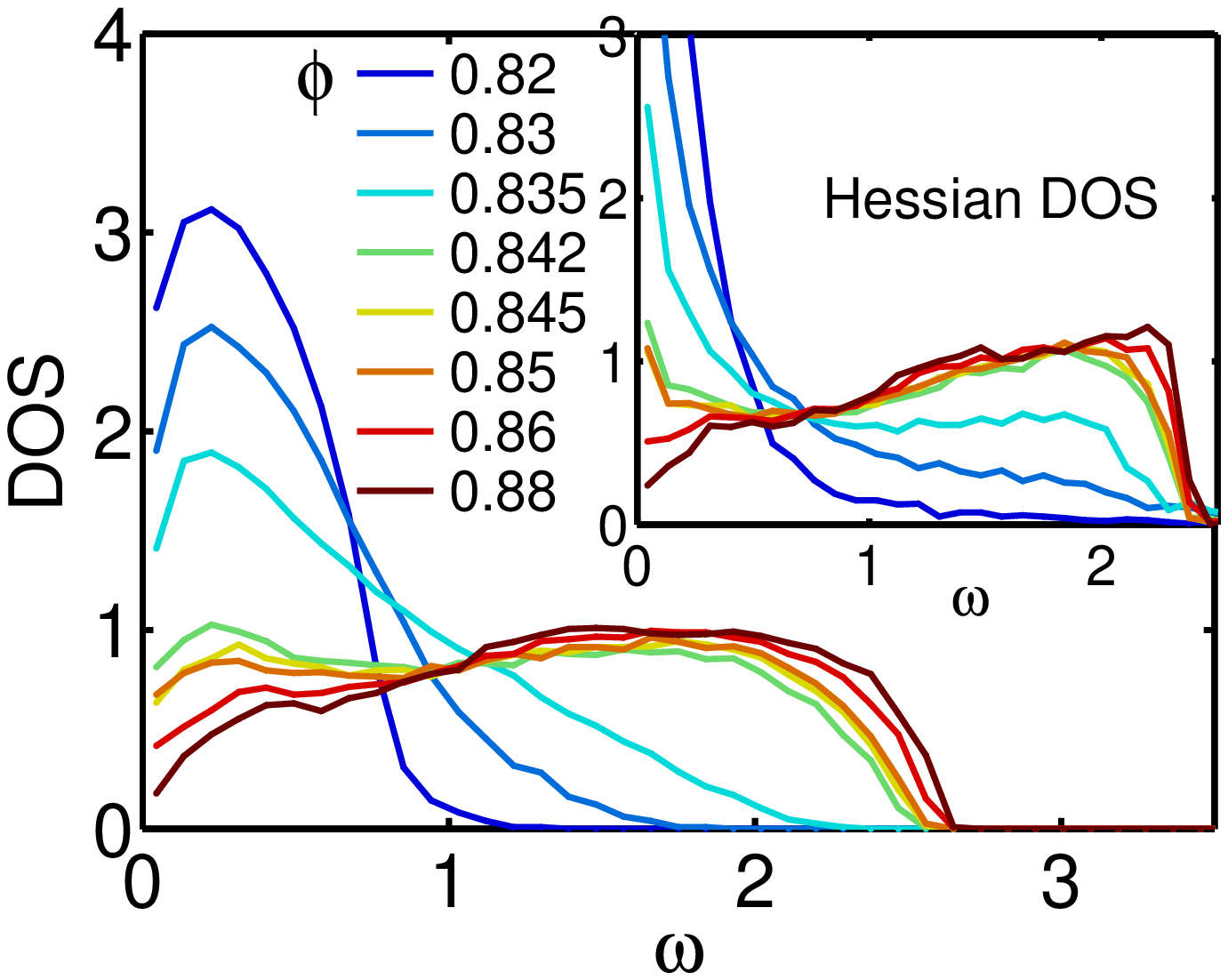}
\caption{ Top Left: Shadow system for a packing of thermal soft sphere just below
jamming at $\phi=0.835$ and $kT=10^{-6}$. 
The contacts are plotted with widths proportional to their effective stiffness
$k_{\text{eff}}$, red with positive $k_{\text{eff}}$ and blue with negative
$k_{\text{eff}}$. Top right: interparticle gap distribution for a range of packing fractions; 
with a negative mean gap above jamming and a positive gap below jamming. 
Bottom left: Dependence of the effective stiffnesses $k_{\text{eff}}$ on the interparticle gap for different $\phi$; the 
different densities sample different regions of the stiffness curve. Bottom right: Density of
states $D(\omega)$ computed from the eigenspectrum of \Cp compared to the one
derived from the Hessian of the soft sphere potential with an added log-potential (inset; see text). All curves for $kT=10^{-6}$. }
\label{fig:shadow_system_SS_NL}
\end{figure}

The differences between the shadow system and the model system 
become strongly apparent if we cross the jamming transition at finite temperature.
Figure ~\ref{fig:shadow_system_SS_NL} shows different aspects of the shadow system
as a function of packing fraction $\phi$ at finite (low) temperature $kT=10^{-6}$.
From the plot of the shadow system at $\phi=0.835$, it is clear that the stiffnesses
now deviate significantly from a step function. The gap distribution (top right) moves
from distributions with only negative gaps (overlaps) above jamming, to a situation
with only positive gaps below jamming. Around jamming, we find both over- and underlaps
within the same packing, and both contribute to the effective stiffness.
Even though the stiffness-gap function has only slightly broadened from a step function, 
the different packing fractions sample different parts of the curve (bottom left), and the
resulting shadow system is significantly different from the model system.
At high densities, we do recover a shadow system density of states which resembles
the Hessian of the soft sphere potential (bottom right, compare to inset).
At lower densities, the similarity breaks down, most clearly below jamming,
where the soft sphere density of states has a large number of zero modes, 
and it is mechanically unstable, unlike the shadow system one.

In the inset to the bottom right figure, we compare the shadow system density of states
to the simplest effective model for thermal soft spheres: $V_{eff}=V_{harm} + kT V_{log}$, that is we have
added an effective logarithmic potential for hard spheres to the soft sphere potential (see below
for a disscussion of the log-potential).
While $V_{eff}$ captures part of the shift of the density of states to lower frequencies below jamming,
it is not a good fit: all the shadow system density of states have a peak at intermediate frequencies,
unlike the density of states obtained through $V_{eff}$.

\begin{figure}
 \centering
\includegraphics[width = 0.75\columnwidth,trim = 30mm 10mm 25mm 5mm,
clip]{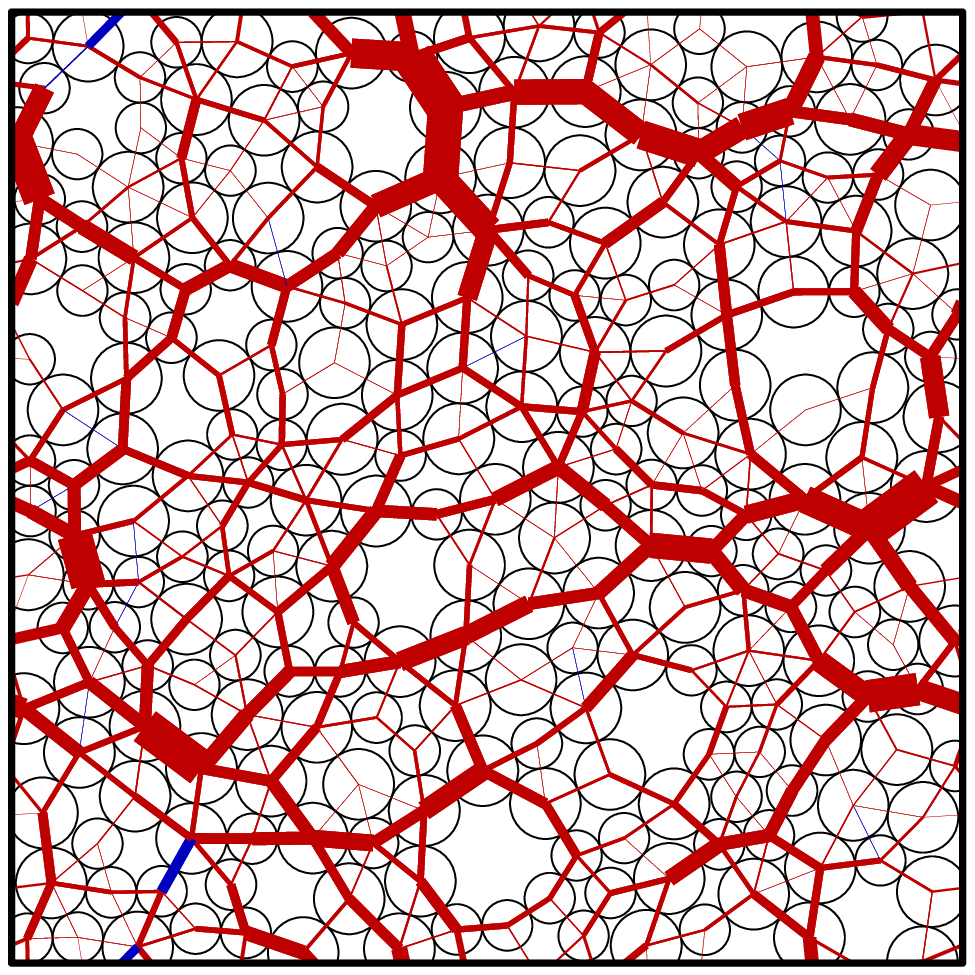}
\includegraphics[width = 0.49\columnwidth, trim = 0mm 0mm 12mm 5mm,
clip]{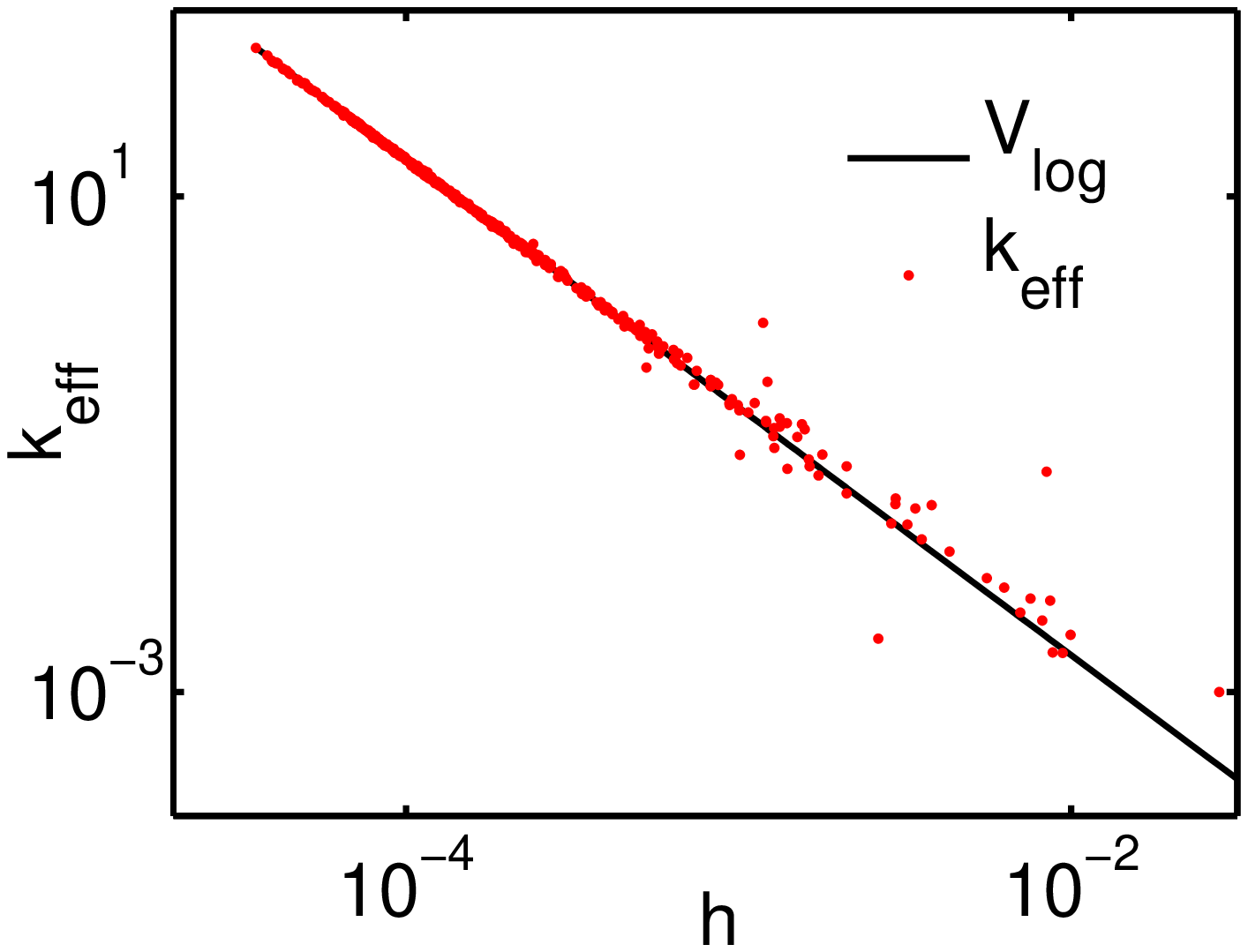}
\includegraphics[width = 0.49\columnwidth,trim = 0mm 0mm 12mm 5mm,
clip]{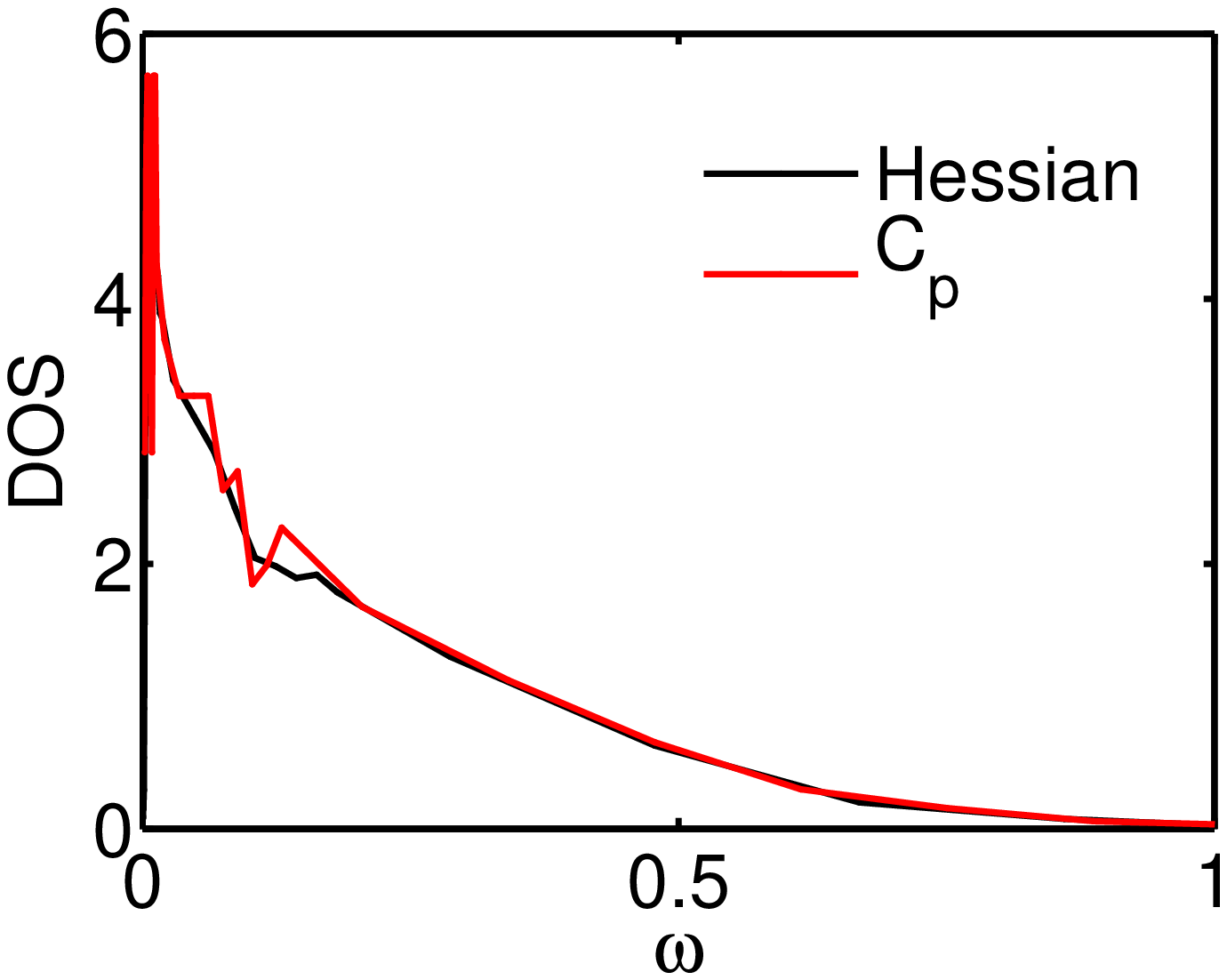}
% Logarithmic version
% \includegraphics[width = 0.49\columnwidth,trim = 0mm 0mm 12mm 5mm,
% clip]{DOS_hard_compare_log.eps}
\caption{Top: Shadow system for a packing of $N=256$ thermal hard spheres just below jamming
 ($\phi =0.83878$, $\phi_J=0.83894$).  The contacts are plotted with widths
proportional to their effective stiffness $k_{\text{eff}}$, red with positive
$k_{\text{eff}}$ and blue with negative $k_{\text{eff}}$. Bottom Left: effective
stiffnesses as a function of the gaps separating the particles: in black the
effective stiffnesses derived from the effective logarithmic potential; in red
those extracted from the inversion of \Cp. Bottom right : Density of states
$D(\omega)$ computed from the eigenspectrum of \Cp compared to the one derived
from the Hessian of the effective potential.}
\label{fig:shadow_system_HS}
\end{figure}

The effective log-potential we just used was first introduced for hard spheres
slightly below jamming where the situation is again not so clear: the
hard spheres potential is singular and, as for the soft spheres case, the system
is a priori not mechanically stable anymore. 
On one hand one can derive an effective pair potential and effective stiffnesses
by computing the forces amongst the particles via the exchange of momentum
during the collisions~\cite{BritoWyart_EPL2006}.
This potential scales like the logarithm of the interparticle gap. On the other hand,
one can compute the effective stiffnesses derived from the inversion of \Cp. 
Figure~\ref{fig:shadow_system_HS} compares both effective stiffnesses for a
system of $N=256$ hard disks just below jamming ($\phi =0.83878$,
$\phi_J=0.83894$).
We indeed find an effective connectivity of the packing even though in their
mean positions, the particles are not touching. This is consistent with the
effective potential approach and we also find $k_{\text{eff}} \sim 1/h^{2}$,
where $h$ is the interparticle gap, consistent with a logarithmic potential.
Hence the shadow system remains a good approximation in this case too. As a
consequence one can again compare the density of state $D(\omega)$ computed from
\Cp and the one derived from the Hessian of the effective potential. One
observes on figure~\ref{fig:shadow_system_HS} that they match pretty well. 
Given the singular character of the hard sphere potential and the fact that the
reference state is only a metastable state, this is a non trivial result which
validates in a self consistent way both the approach in terms of an effective
potential introduced in~\cite{BritoWyart_EPL2006} and the use of \Cp as a tool
for extracting the density of states for slightly unjammed systems.

Before closing this section, let us mention that we have followed the same
analysis in the case of a real experimental system composed of NIPA colloidal
particles above jamming~\cite{PRL_NIPA_2010}. Conceptually, it is very similar
to the soft sphere simulation we have just discussed. However, as already
mentioned, the presence of a solvent and electrostatic interactions could alter
the description in terms of a simple repulsive pair potential. We found that the
effective stiffness is far above the noise for adjacent particles only. This
tells us that elastic interactions between neighboring particles dominate the
data, as they should. Furthermore, the effective stiffnesses as a function of
the interparticle separation could be measured and is compatible with an
harmonic or Herzian interaction. Hence also here the shadow system captures the
essence of the vibrational properties of the real system under consideration.
The reader interested by this experimental study should refer to the original
paper.

\subsection{Principal component analysis}
%*******************************************************************************
\label{PCA}

So far we have discussed how to interpret the eigenmodes and eigenvalues of \Cp
for thermally equilibrated systems. We have seen in section~\ref{non-equilibrium}
that for non equilibrium systems, the distribution of the energy on
each mode must be known before extracting the density of state. This in general
is not the case. However the spectrum of \Cp and the associated modes still
convey a lot of information about the dominant modes of relaxations.

Such a path has been followed in analyzing different systems for which the
underlying microscopic dynamics is unknown. In~\cite{Laloux-PRL1999}, 
the price fluctuations in financial markets are analyzed by comparing the spectrum of
eigenvalues of the covariance matrix with the results of random matrix theory.
The deviations from the purely random case contains the true information about
the prices. In \cite{Micheletti-PCA2004} the protein near-native motion is
characterized. Although molecular dynamics simulations based on all-atom
potential allows the identification of a protein functional motion on a wide
range of timescales,  the very large time scales are not easily accessible in
simulations. A way to bypass this problem is to study the correlations of the
fluctuations of the deviation of the protein backbone from its native
configuration. The principal components of this matrix correspond to the
collective modes of the protein which happen on larger time scales.

Two of the present authors have used a similar approach in the case of vibrated
granular media~\cite{Brito_SM2010}.
The  experimental  system consists in a 1:1 bidisperse monolayer of brass
cylinders. Mechanical energy is injected in the bulk of the system by horizontally vibrating
 the glass plate on which the grains stand and dissipated by
solid friction.   The experimental protocol produces very dense steady states with a
packing fraction up to $\phi = 0.8457$. The stroboscopic motion of a set of
$1500$ grains is tracked in the center of the sample. The set-up, the quench
protocols and the main properties of the system, are described in detail in
\cite{LechenaultEPL1, LechenaultEPL2}.

\begin{figure}
\centering
\includegraphics[width = 0.85\columnwidth, clip,angle=-90]{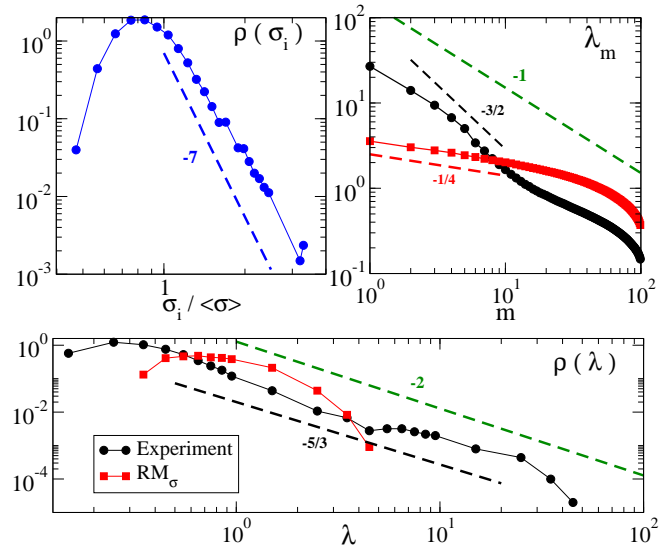}
\caption{Eigenvalues of \Cp for jammed grains under horizontal vibration above
jamming. Top-left : Probability density of the rms displacement of
each grain around it average position in the metastable state. 
Top-right : positive eigenvalues sorted in decreasing order.
Bottom : probability density of the positive eigenvalues. On the
center and right plots the black dots  correspond to the real dynamics and the
red square to the $RM_\sigma$ model;  the dotted line correspond to the expected
distribution for a crystal according to the Debye law}.
\label{spectrumPropertiesGrains}
\end{figure}

We now briefly illustrate the methodology adapted in~\cite{Brito_SM2010} to
discuss the spectral properties of \Cp.  We have considered a subset
of $N=350$ grains acquired during $T=100$ time steps at a packing fraction
$\Phi=0.844$ so that the metastable state of reference is well defined
(no rattler, no structural rearrangement). 
Because $r=dN/T$, where $d$ is the space dimension, is larger than one, 
strictly zero eigenvalues appear in the spectrum of \Cp.
We shall come back to this practical issue in the next section and restrict the
discussion to the positive eigenvalues. The fluctuations of each
grain $i$ around its metastable position are Gaussian but the width, $\sigma_i$,  of
the distribution is widely distributed amongst the particles. according to
the fat tail distribution $\rho(\sigma)\sim \sigma^{-(1+\mu)}$ plotted on 
figure~\ref{spectrumPropertiesGrains}-top-left , where $\mu \approx 6$.

The Mar\u{c}enko-Pastur theorem, the purpose of which is to relate the
spectrum computed at finite $r$ to the one of the true correlation matrix
could be used to obtain predictions but no explicit analytic form
for the shape of the eigenvalue spectrum of \Cp. Please
see appendix ~\ref{appendix_MP} for details. One alternative strategy
is to define a Random Model of \emph{uncorrelated} Gaussian variables
the width $\sigma_i$ of which are the experimental ones. We will refer to this
model as the $RM_\sigma$ case. The eigenmodes extracted from the real
dynamics with eigenvalues larger than the largest one of the $RM_\sigma$ model
contain the relevant information about the correlations. 

Figure~\ref{spectrumPropertiesGrains}-top-right, respectively
figure~\ref{spectrumPropertiesGrains}-bottom, displays the eigenvalues $\lambda_m$
sorted in decreasing order, respectively the probability density $\rho(\lambda)$
extracted from the real dynamics as compared to those obtained for the
$RM_\sigma$ model. Note that there is a straightforward correspondence between
these two curves since $\frac{1}{N}m(\lambda)$ is exactly the inverse cumulated
distribution of $\lambda$ and a power-law behavior $\lambda_m \sim
m^{-1/\alpha}$ translates into a density $\rho(\lambda)\sim
\lambda^{-(1+\alpha)}$.  The five largest eigenvalues for the experimental system are clearly larger than
the largest one obtained for the $RM_\sigma$ model.
This comparison shows unambiguously that the top eigenvalues of \Cp contain
useful information about the dynamics of the system, and are not drowned in
noise. It also demonstrates the existence of strong spatial correlations: by
moving together, particles achieve large collective fluctuations that would not
develop otherwise.

If one should not try to convert these eigenspectrum into density of states,
it is however always possible to convert the expected density of states of a thermal crystal
into the the spectrum of its dynamical matrix. Here, a 2d crystal, with a density of states
$\rho(\omega) \sim \omega^{d-1} = \omega$, with $\omega \propto \lambda^{1/2}$,
has $\rho(\lambda)\sim \lambda^{-2}$, that is $\alpha = 1$ and $\lambda_m \sim 
m^{-1}$ as indicated on the figures by the green dotted line.  
The experimental data are consistent with an estimated
$\alpha\approx 2/3$,  that is a slower decay of the spectrum than for the
crystal case. One can interpret this result saying that there is a larger
fraction of modes participating in the dynamics than in the crystalline case.

One can also analyze the structure of the most significant modes. This was done
in~\cite{Brito_SM2010} for this system of bi-disperse grains when approaching
jamming from above, and it could be demonstrated quantitatively that there is a
redistribution of spectral weight towards larger eigenvalues corresponding to
softer modes and that the softest mode becomes more coherent and spatially
organized.  However, as it was discussed above, for such a system where the
energy is not distributed evenly over the modes, one cannot convert this information into the
vibrational properties and the density of states of the system.  The results presented
here as well as in~\cite{Brito_SM2010} illustrate well how the PCA of the dynamics
remains a fruitful tool of analysis without referring to the vibrational properties.

%*******************************************************************************
\section{Practical guideline}
%*******************************************************************************
\label{practik}

In the previous section, we have assumed that the data at hands were \lq\lq
perfect\rq\rq in the sense that neither statistical limitations nor resolution
issues came into play. Also we did not consider the possibility of a strong
anharmonicity of the dynamics and we have not discussed the way, one should get
rid of the rattlers, those particles which do not participate in the collective
motion but instead ``rattle''  in their cages.

Measuring \Cp to a sufficient level of precision to extract reliable information
from it is not a simple affair. The goal of this section is to disentangle
all the possible sources of practical problems that one faces when computing \Cp
and to propose methods and alternatives to obtain a faithful information from
its spectrum.

In subsection~\ref{test_metastability}, we start by discussing a way to test
the metastability of the reference state. Indeed all the approach is based on
the fact the displacement field  is  purely composed of vibration around a
steady reference state. We illustrate this issue in the case of hard spheres,
which by definition sit below jamming, where metastable states only have a
finite lifetime. It is thus of crucial importance to control that the system
does not escape the metastable state during the time window of the analysis. 
In subsection~\ref{harmonicity}, we then illustrate on the case of vibrated
granular media, how one can observe anhamonicity setting in when approaching the
jamming transition from above. The reader may also refer to~\cite{O-HernPRL}, where the
same issue is discussed in the case of numerical simulations of soft spheres.
As an immediate consequence of the above effects, the time window of the
analysis may be seriously shortened as compared to the full duration of the
numerical or real experiment.  Hence statistical issues come into play. 
More specifically, we shall see in sections~\ref{test_stat} and ~\ref{test_conv}
how to deal with statistical independence of the data as well as convergence issues.  
In the case where the data sets have been obtained
experimentally, resolution issues may also alter the analysis.  We thus consider
in subsection~\ref{exp_res} a NIPA colloidal experiments to illustrate how
convergence and experimental resolution interplay to affect the computation of
the density of states. 
Finally, identifying rattlers is usually a challenging matter. It emerges that
calculating the eigenspectrum itself is a good ``filter'' for rattlers:  their
motion appears only in a few low energy eigenvalues which do not mix with the
remainder of the motion. This is discussed in section~\ref{rattlers}.

\subsection{Metastability}
%*******************************************************************************
\label{test_metastability}

Before starting any of the analysis discussed in the previous section, one must
have checked that the dynamics is purely composed of fluctuations around a well
defined reference state. This can be done both a priori, in order to select a
good time window to perform the analysis, and a posteriori on the basis of the
properties of \Cp.  A natural tool to characterize the dynamics is the
self-intermediate correlation function, defined as
\begin{equation}
 C_q(t_0,t) = \langle~ \cos (\vec q ~.\Delta \vec r_i(t,t_0) ~\rangle _j,
\label{Cq}
\end{equation}
\noindent
where the average is made on the particles but not on the initial time. $\Delta
\vec r_i(t,t_0)=\vec r_j (t) - \vec r_j(t_0)$ is the displacement of the
particle $j$ between $t_0$ and $t+t_0$, $\vec q$ is a vector whose amplitude is
given by $q=\pi/a$ and $a$ is a length scale which sets the amplitude
of the displacement above which the particle motion induces a change of
metastable state. There is obviously a large part of arbitrariness in such a
definition and traditionally $a$ is set to a fraction of the particles diameter, 
so that a metastable state more or less corresponds to a given configuration
of neighbors.

\begin{figure}
 \centering
\includegraphics[width = 0.8\columnwidth, clip,angle=-90]{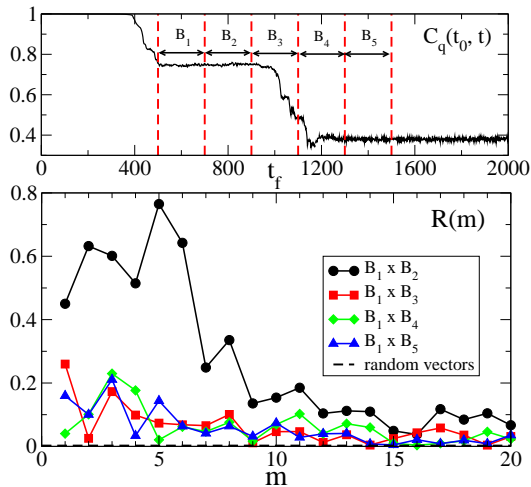}
\caption{Probing metastability. Top : $C_q(t_0,t)$ for a system of hard spheres
below jamming ($N=1024$ and $\phi = 0.841947$; $\phi_J=0.841959$). The time unit
is $t_f$, the time interval at which particle positions are stored, and
corresponds to $800$ collisions per particle. Each interval delimited by dashed
lines has $200$ snapshots, during which we compute \Cp and its  basis $B_i$ of
eigenvectors  $\{ |\lambda_{m} \rangle \}$.
Bottom : Robustness of the first $20$ modes as defined in Eq.(\ref{Rm}) between 
pairs of intervals $B_i \times B_j$ as indicated in the legend. The dashed line
indicates the value of $R(m)$ for purely random eigenvectors. }
\label{Cq_MSD}
\end{figure}

Figure~\ref{Cq_MSD}-top displays $C_q(t_0,t)$ computed for a system of $N=1024$
bidisperse hard discs the dynamics of which has been simulated
using molecular dynamics slightly below jamming ($\phi = 0.841947$;
$\phi_J=0.841959$)~ \cite{BritoWyart_EPL2006, BritoWyart_JSTAT2007}. Here $a$
has been set to the  radius  of the smaller particles.
One observes very well defined plateaus, during which the system is trapped in a
metastable state, separated by a quick relaxation event that we call {\it
crack}. Typically, the size of the plateau defines $T$, the interval of time
available to compute \Cp. There are however some cases where one cannot identify
the plateaus so easily. As a matter of fact this will happen each time the
system size is large as compared to the typical size of a crack. When
this is the case, one can always cut the system into smaller subsystems and do
the analysis in each subsystem as illustrated in~\cite{Brito_SM2010}.

As long as the system remains in the same state of reference, the basis of
eigenvectors should essentially remain unchanged. This can be checked by
dividing the time interval of duration $T$ in smaller subintervals and computing
an indicator of the modes robustness, defined as \cite{Brito_SM2010}:
\be
R(m) = \sum_{j=-M}^{j=+M} \langle \lambda_m |  \lambda'_{m+j} \rangle^2, 
\label{Rm}
\ee
\noindent
where $\langle \lambda_m |  \lambda'_{m +j} \rangle$ is the scalar product
between the modes computed during two successive observation windows of duration
$T'<T$. Setting $M$ to a small but strictly positive value allows neighboring
modes to possibly exchange their rank. As long as the largest eigenvalues are
concerned, we could check that the results are basically independent of $M$ for
$M \ge 2$ and we fixed $M=2$. Figure~\ref{Cq_MSD}-bottom displays $R(m)$
for the twenty largest eigenvalues computed between intervals defined as indicated on
figure~\ref{Cq_MSD}-top by vertical dashed lines. The robustness of the the
first modes is higher when computed amongst intervals belonging to the same
plateau. As soon as one compute the modes on an interval including the crack,
the robustness decreases sharply and rapidly reaches the baseline level
$R(m)=(2M+1)/2N$, one would obtained for purely random vectors. Note also that
the robustness is not restored when the two intervals belong to different
plateaus indicating that the system has indeed relaxed from one metastable state
to another.

\subsection{Harmonicity of the dynamics}
%*******************************************************************************
 \label{harmonicity} 

In the previous section, we have identified a metastable state and checked a
posteriori that the basis of eigenvectors remains unchanged during the lifetime
of this state. As a matter of fact, one must also check that the dynamics around
this state is purely harmonic. For an equilibrium system close to jamming, the
structure of the packing is frozen and fluctuations of the particles are
thermal, so that assuming Gaussian fluctuations around the metastable state
sounds reasonable. However  it was recently claimed that repulsive contact
interactions make jammed particles systems inherently anharmonic~\cite{O-HernPRL}.
The authors argue that  at very low temperature the breaking and forming of
inter-particles contact is responsible for anharmonicity in the sense that the
response does not remain confine to the original mode of excitation. Such a
definition of harmonicity is very strict, and how the result depends on system
size, temperature and distance to jamming is still a matter of debate. It is clear
 that for mechanically excited athermal and dissipative systems, such as shaken
macroscopic grains, there is no reason why the dynamics should be
harmonic.

\begin{figure}[t!]
\begin{center}
\rotatebox{-90}{\resizebox{85mm}{85mm}{\includegraphics[trim = 0mm 0mm 0mm
85mm, clip]{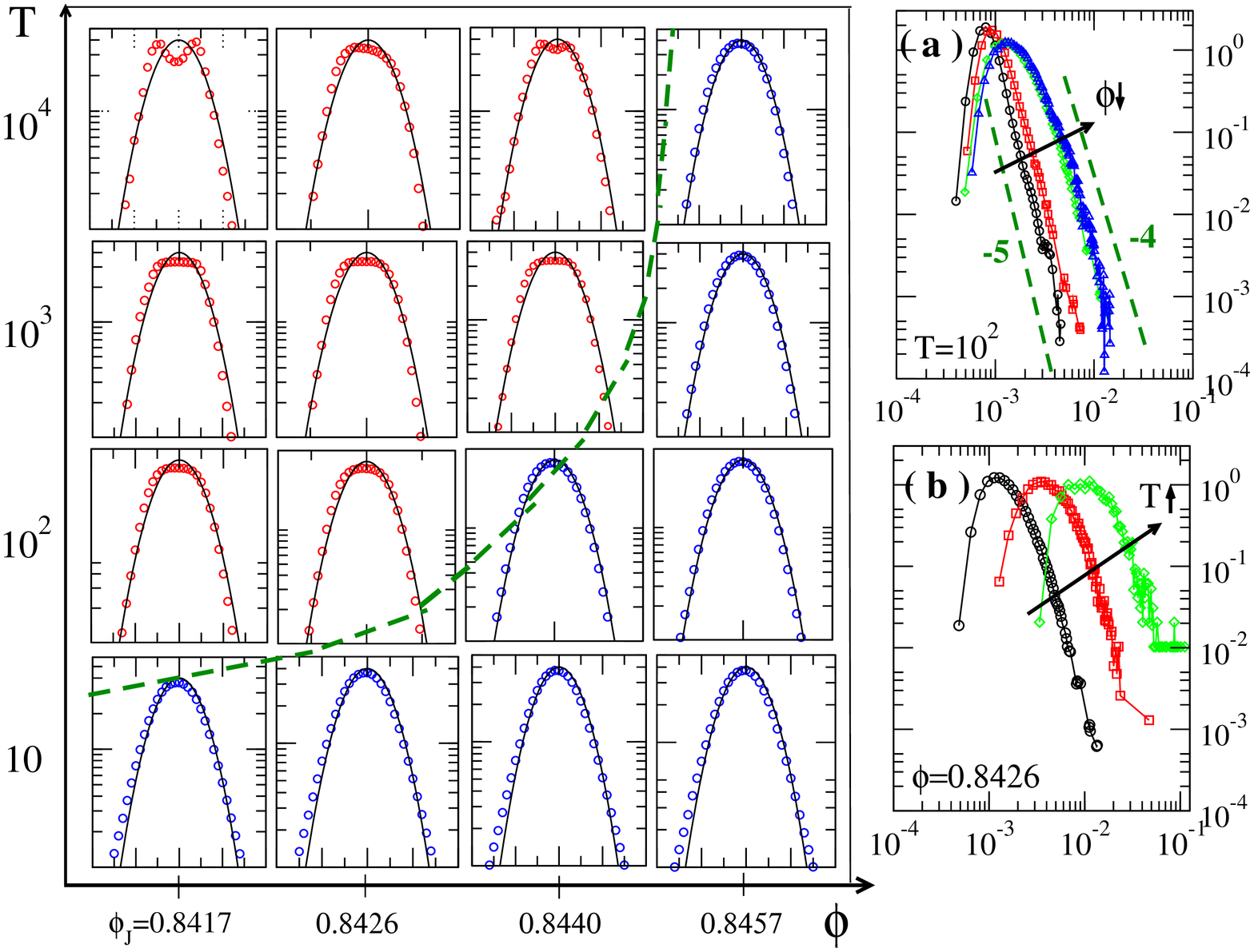}}}
\caption{Distributions of the position fluctuations $\rho(\delta x_i/\sigma_i)$
for four values of the packing fraction and four durations of observation $T$. 
We focus on the top of the distribution, which is compared to a Gaussian
(continuous black line). The green hatched line separates the distributions
well described by a Gaussian from those deviating from this harmonic description }
\vspace{-0mm}
\label{diagram}
\end{center}
\end{figure}

While studying horizontally shaken grains, two of the present authors proposed
to check the harmonicity of the dynamics in a naive but experimentally
accessible way, namely by computing the distribution of the \emph{individual}
fluctuations: $\rho(\delta x_i/\sigma_i)$, where $\delta x_i$ is the position
fluctuation along a given direction and $\sigma_i$ is the root mean square
displacement of particle $i$.
Figure~\ref{diagram} presents such distributions obtained for a system of
$N=1550$ brass disks shaken horizontally on a oscillating plate just above
jamming ($\phi_J=0.8417$, see~\cite{LechenaultEPL1} for details on the
experimental set up and protocol).   The distributions are computed for four
different packing fractions $\phi$ and four durations $T$ of the window of
observation. They are then ensemble averaged over the $10^4/T$ intervals
provided by the full dataset. 

The distributions shown in figure~\ref{diagram} highlight some important
characteristics of the dynamics. The parameter space $(\phi,T)$ can be divided
into two regions, as illustrated by the green hatched line: for small enough
observation duration $T$ or large enough packing fractions, the distributions
are unimodal with a Gaussian core: particles jiggle around a well defined
average position; for longer $T$ or smaller densities, the distribution starts
developing a flat top, with a poorly defined maximum. This suggest that on these
longer observation times, the average position of a significant part of the
particles is not well defined anymore. Particles either drift slowly or even
find (collectively) another metastable position, as suggested by the double peak
observed in the case $\phi=0.8417$ and $T=10^4$, i.e. for the loosest packing
fraction and the longest observation time. This means that over long time
scales, the evolution of the average position becomes comparable or even larger
than the fluctuations, and it becomes meaningless to describe the system in
terms of small vibrations around a fixed metastable state. For an infinite size
system, some rearrangement always happens somewhere, and the covariance matrix
${\bf C}_p$ is always ill-defined. The ``allowed'' time scale $T_{\max}(\phi,L)$
is expected to scale inversely with the system size $L$. 

\subsection{Statistical independence}
%*******************************************************************************
\label{test_stat}

Once a reference state is identified during a time window of duration $T$, one
needs to know whether the particle positions stored on successive snapshots are
independent, or at least sufficiently uncorrelated to compute \Cp. Again
computing the robustness of the basis of eigenmodes is a good way to check the
validity of the analysis: if the particle positions are not independent the
correlator \Cp is badly estimated and the eigenmodes are mostly composed of
noise. To illustrate this point we use again the hard discs system just below
jamming ($\phi = 0.83878$; $\phi_J=0.838865$), now with $N=256$, for which we
could identify a long lasting metastable state the total duration of which, $T$,
corresponds to $1.6 \times 10^6$ collisions per particles.

\begin{figure}
\centering
\includegraphics[width = 0.84\columnwidth, clip,angle=-90]{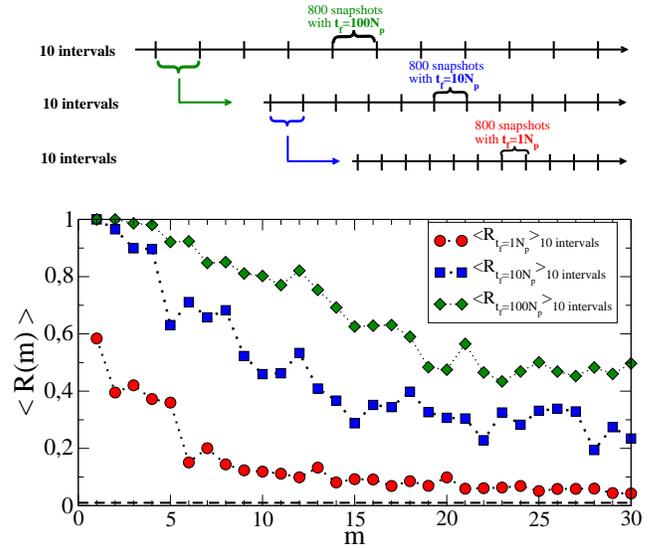}
\caption{Testing statistical independence during a given metastable state for a system
of $N=256$ hard disks below jamming ($\phi =0.83878$; $\phi_J=0.838865$). 
Top : sketch of the time discretization procedure (see text for details). 
Bottom : $\langle R(m) \rangle$ for the three discretization scheme as indicated by the legend.}
 \label{StatIndep}
\end{figure}

Figure~\ref{StatIndep}-top sketches the procedure. The total acquisition is
divided in $10$ intervals. Each of them contains $800$ snapshots of the
positions, separated by $t_f=200$ collisions per particle. \Cp is computed and
diagonalized on each interval, and the average robustness among successive
intervals $\langle R(m) \rangle$ is evaluated for the first $20$ largest
eigenvalues. To preserve the same statistics, together with a reduced interval
of time between the snapshots, we then subdivide each interval in $10$
subintervals, each of them containing again $800$ snapshots of the positions,
but now separated by $t_f=20$ collisions per particle. And the procedure is
iterated once more leading to $800$ snapshots of the positions separated by
$t_f=2$ collisions per particle in each interval.
Figure~\ref{StatIndep}-bottom compares $\langle R(m) \rangle$ for the three
successive level of time discretization. Although the system remains in the same
metastable state, the number of snapshots and the averaging used to compute \Cp
are identical in all three cases, one clearly observes the convergence of
$\langle R(m) \rangle$ towards larger values when the snapshots are well
separated. In the present case the finest discretization is obviously irrelevant
and a separation of successive snapshots larger then about $50$ collisions per
particle is necessary.

Altogether, whether it comes from experimental limitations of the acquisition
rate or from a microscopic timescale inherent to the dynamics, there is always a
minimal timescale separating the accessible and useful snapshots of the particle
positions. Since we have also seen that the analysis is confined to the lifetime
of the reference state, we are now in the position to face the convergence issue
of the eigenspectrum as discussed by the Mar\u{c}enko Pastur theorem.

\subsection{Convergence of the spectrum (small sample limit)}
%*******************************************************************************
\label{test_conv}

The Mar\u{c}enko-Pastur (M-P) theorem~\cite{MarcenkoPastur, Silverstein,ElKaroui, Laloux-PRL1999}
 applies to the ensemble of Wishart random matrices;  that is
the ensemble of $p\times p$-matrices constructed by the addition of $n$
\emph{independent} samples $x_{i} x_{j}$, where the $x_{i}$ ($i \in 1 ... p$) are distributed
according to either a multivariate Gaussian distribution or a distribution with
finite higher moments, similar to the conditions of the central limit theorem.
The M-P theorem then predicts the eigenvalue distribution of such matrices in
the limit $n\rightarrow \infty$, $p\rightarrow \infty$, $ r = p/n$ finite. However,
as one can see in appendix~\ref{appendix_MP}, there is no explicit form
for the distribution.  Also, the theorem tells nothing about the convergence of
the eigenmodes.  This is why we concentrate here on the numerical characterization
of the convergence in a specific and well controlled situation. This will allow us to
discuss both the shape of the spectrum and the relevance of the modes themselves.
Also, it will allow us to examine the case where $r>1$, when a finite part of the
spectrum is strictly zero. The reader interested in analytical predictions may
refer to the original papers ~\cite{MarcenkoPastur, Silverstein, ElKaroui, Laloux-PRL1999},
as well as to appendix~\ref{appendix_MP}.

\begin{figure}
\centering
\includegraphics[width = 0.80\columnwidth, clip,angle=-90]{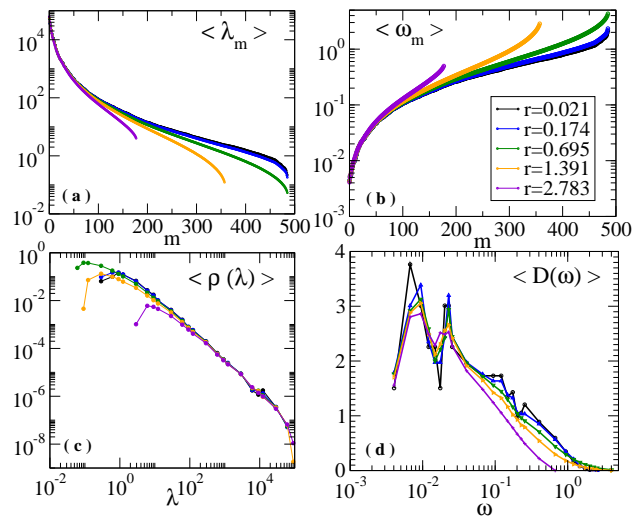}
\caption{Convergence of the spectrum of eigenvalues of \Cp for a system of
$N=256$ hard disks below jamming ($\phi = 0.83878$; $\phi_J=0.838865$) for
decreasing values of $r=2N/n$ as indicated in the legend. Top: Eigenvalues (left)
and corresponding eigenfrequencies (right) sorted in decreasing,
respectively increasing order.  Bottom: Eigenvalues spectra (left) and
corresponding density of states (right).}
\label{fig:convspect}
\end{figure}

We consider the same system as above, with $N=256$ hard disks just below jamming
($\phi = 0.83878$; $\phi_J=0.838865$) within a long lasting metastable state of
total duration $T=4.6\,10^6$ collisions per particles and we choose to retain as
independent particle positions separated by $200$ collisions per particle, so
that $n=23000$.  In section~\ref{shadow}, we have shown that the spectrum of \Cp
and that of the effective dynamical matrix match when the whole set of data
is used to compute \Cp. Here we discuss how much the spectrum and the modes of
\Cp deviate from their asymptotic behavior when the time interval used to
compute \Cp is artificially reduced such that $r=2N/n$ varies in the range
$[0.021-2.28]$. 

\begin{figure}
\centering
\includegraphics[trim = 30mm 0mm 20mm 0mm, width = 0.5\columnwidth, 
clip,angle=-90]{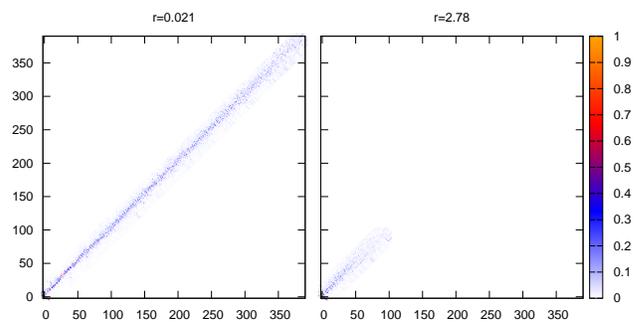}
\caption{Matrices of the scalar product between the modes of  \Cp  and the
modes of  the dynamical matrix for a system of $N=256$ hard disks below
jamming ($\phi = 0.83878$; $\phi_J=0.838865$). The rows are the indices
of the modes in increasing order of their frequency and the colorbar indicates
the value of the scalar product. Left: $r=2N/n=0.021$  Right: $r=2N/n=2.78$.}
\label{fig:convmodes}
\end{figure}

Figure~\ref{fig:convspect} displays the spectral properties of \Cp for
increasing values of $r$.  As soon as $r>0.5$, the smallest
eigenvalues are seriously underestimated, respectively the largest frequencies
are overestimated. For $r>1$, strictly zero eigenvalues replace the lowest
eigenvalues. On the contrary, the largest eigenvalues, respectively the lowest
eigenfrequencies, are surprisingly robust, even in the cases $r>1$.
Because of this robustness, when one is interested in the low
frequency part of the density of states, it is recommended to normalize the
density such that the area under the curve equals the fraction of non-zero
eigenvalues. As a result, the density of states  $D(\omega)$ conserves its shape
as long as $r<1.5$ and only for larger $r$ the large frequencies part of the
density of states starts to be significantly depleted. 

Finally figure~\ref{fig:convmodes} illustrates the robustness of the modes
themselves. To do so, we define the matrix of scalar product between the two
eigenbasis: the modes of  \Cp and  the modes of the effective  dynamical matrix.
If the modes were strictly the same, this matrix should be the identity matrix.
Figure~\ref{fig:convmodes}-left shows the matrix of the scalar products for
$r=0.021$, from where we observe that the modes of \Cp project on a very
small number of modes of the effective dynamical matrix, as already
pointed at in section~\ref{shadow}. We see that even for $r=2.78$, the modes of \Cp
associated with the largest eigenvalues still project on a small number of modes of
the effective dynamical matrix, while the modes of \Cp with a higher frequency are
spread across the whole spectrum of the effective dynamical matrix.

\begin{figure}
\centering
\includegraphics[width = 0.85\columnwidth, clip,angle=-90]{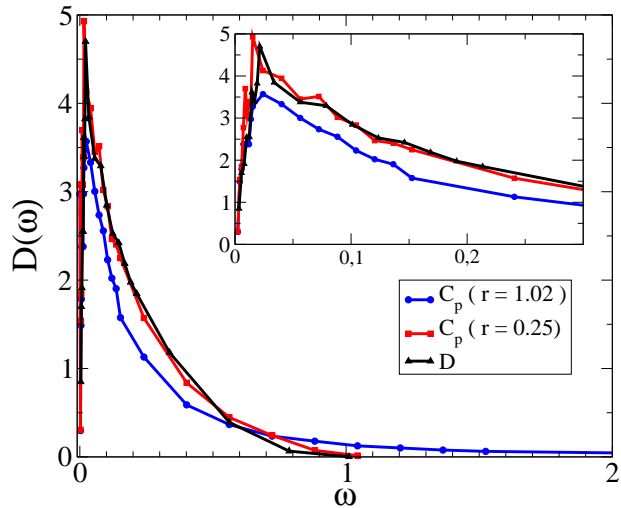}
\caption{Density of states for a system of hard disks below jamming 
($\phi =0.841947$; $\phi_J=0.841959$)  extracted from \Cp either computed using the whole
system of $N=1024$ hard disks (in blue) or cutting the system in 4 sub-systems with 
$N_p^{sub} = 256$ each (in red) as compared to the density of states obtained
from the effective dynamical matrix computed for the whole system (in black).
Inset: zoom on the lowest frequency part of the density of states.}
\label{Cp_M_Dw}
\end{figure}

Altogether, as long as $r<1.5$, the most significant eigenmodes and eigenvalues
of the dynamic matrix as well as the qualitative features of the density of
states are well captured by the analysis of \Cp. Practically speaking, when very
long lasting metastable states are available it is thus often advantageous to
average the spectrum or the density of states on successive time windows,
provided that the duration of each of them satisfies $r\approx 1$.  

\begin{figure}
\centering
\includegraphics[trim = 30mm 0mm 20mm 0mm, width = 0.5\columnwidth, 
clip,angle=-90]{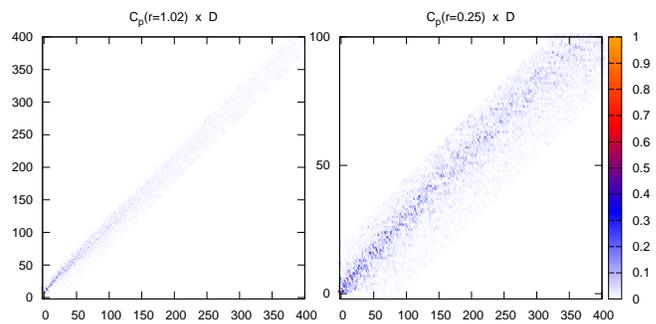}

\caption{Matrices of the scalar products between the modes of \Cp and the
modes of the dynamical matrix for a system of $N=1024$  hard disks below
jamming. Left: Modes of \Cp computed on the whole system. Right: Modes
of \Cp reconstructed from the modes computed separately in four
sub-systems of size $N_p^sub=256$. The horizontal (resp. the vertical) indices label
the modes of the dynamical (reps. the covariance) matrix. The color bar
indicates the value of the scalar product.}
\label{Cp_M_SP}
\end{figure}

When there is no enough snapshots to recover the whole spectrum of modes, 
an alternative strategy is to reduce the system size in order to increase $r$.
In principle one can divide the system in smaller subsystems, and compute
the density of states in each of them. This strategy allows to better access
the low frequency part of the density of states, although the lowest accessible
frequency increases as the inverse of the system size.
Figure~\ref{Cp_M_Dw}  and figure~\ref{Cp_M_SP} illustrate the application of
such a strategy to a system of $N=1024$ particles below jamming ($\phi =
0.841947$; $\phi_J=0.841959$). In this example, the  metastable state has a
total duration of $T=1.6\times 10^5$ collisions per particles and we store 
independent particle positions separated by $80$ collisions per particle. This
means that $n=2000$ snapshots and  $r=1.02$. From  figure~\ref{fig:convspect} 
we know that for this value of $r$ the large frequencies are underestimated.
By cutting the system in 4 boxes, each one has $r=0.25$ and we are now able to
recover the whole spectrum. In figure~\ref{Cp_M_Dw} we compare the density
of states $D(\omega)$ extracted from \Cp either using the whole system of
$N=1024$ hard disks $(r=1.02)$ or cutting the system in 4 sub-systems with 
$N_p^{sub} = 256$ each $(r=0.25)$ to the density of states obtained
from the effective dynamical matrix computed for the whole system.
Whereas the spectrum obtained from \Cp computed with the whole system
only captures the very low frequencies, the one computed for the sub-systems
matches the spectrum from the dynamical matrix on the whole range of
frequencies.

Concerning the robustness of the modes, once the modes have been
computed in each subsystem, we first reconstruct artificial modes for the
whole system by pasting together the modes of each subsystem. Note that the
number of modes in these 2 basis is not the same: for the whole system there are
$2N_p = 2048$ modes while in each subsystem has $2N_p^{sub} = 512$ modes.
Figure~\ref{Cp_M_SP} shows the matrices of the scalar products between the
modes of the dynamical matrix and those of \Cp, either computed on the
whole system (left) or reconstructed as described above from the modes
computed on the subsystems (right). Despite the artificial nature of the modes
obtained from the computation of \Cp in each subsystem, their projection on the
modes of the dynamical matrix remains rather concentrated, especially when
one remembers that each mode of \Cp in this case projects approximately on
four modes of the dynamical matrix.

Let us end this section by quoting a recent work~\cite{Schindler-condmat1111.1419}
which shows that in general the modes are modified by the truncation of
the full real-space correlations and proposes windowing strategies to best avoid
these boundary conditions artifacts. Implementing such strategies on top of the
segmentation proposed here would certainly deserve proper attention.

\subsection{Convergence and experimental resolution (large sample limit)}
%*******************************************************************************
\label{exp_res}

Even in the most favorable case of all experimental situations discussed above,
we have not yet addressed the issue of the experimental resolution. As detailed in appendix~\ref{appendix_MP}, 
the Mar\u{c}enko-Pastur theorem allows us to
compute the impact of a finite resolution on the spectral properties of \Cp at a
finite $r=2N/n$, where $n$ is again the number of independent snapshots of the
particle positions.

\begin{figure*}[t]
 \centering
\includegraphics[width = 0.7\columnwidth,trim = 0mm 0mm 10mm 5mm,
clip]{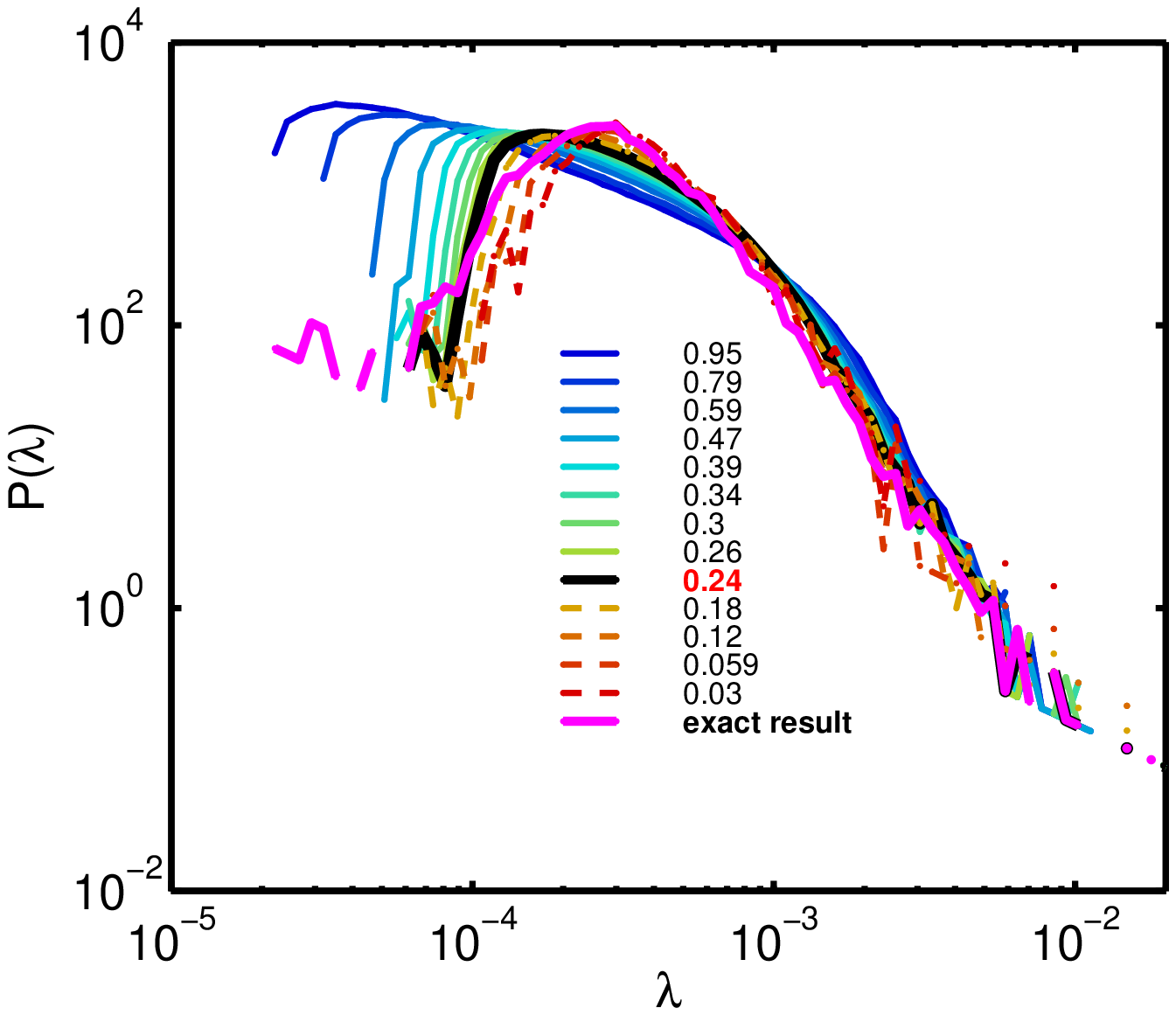}
\includegraphics[width = 0.75\columnwidth,trim = 0mm 0mm 10mm 5mm,
clip]{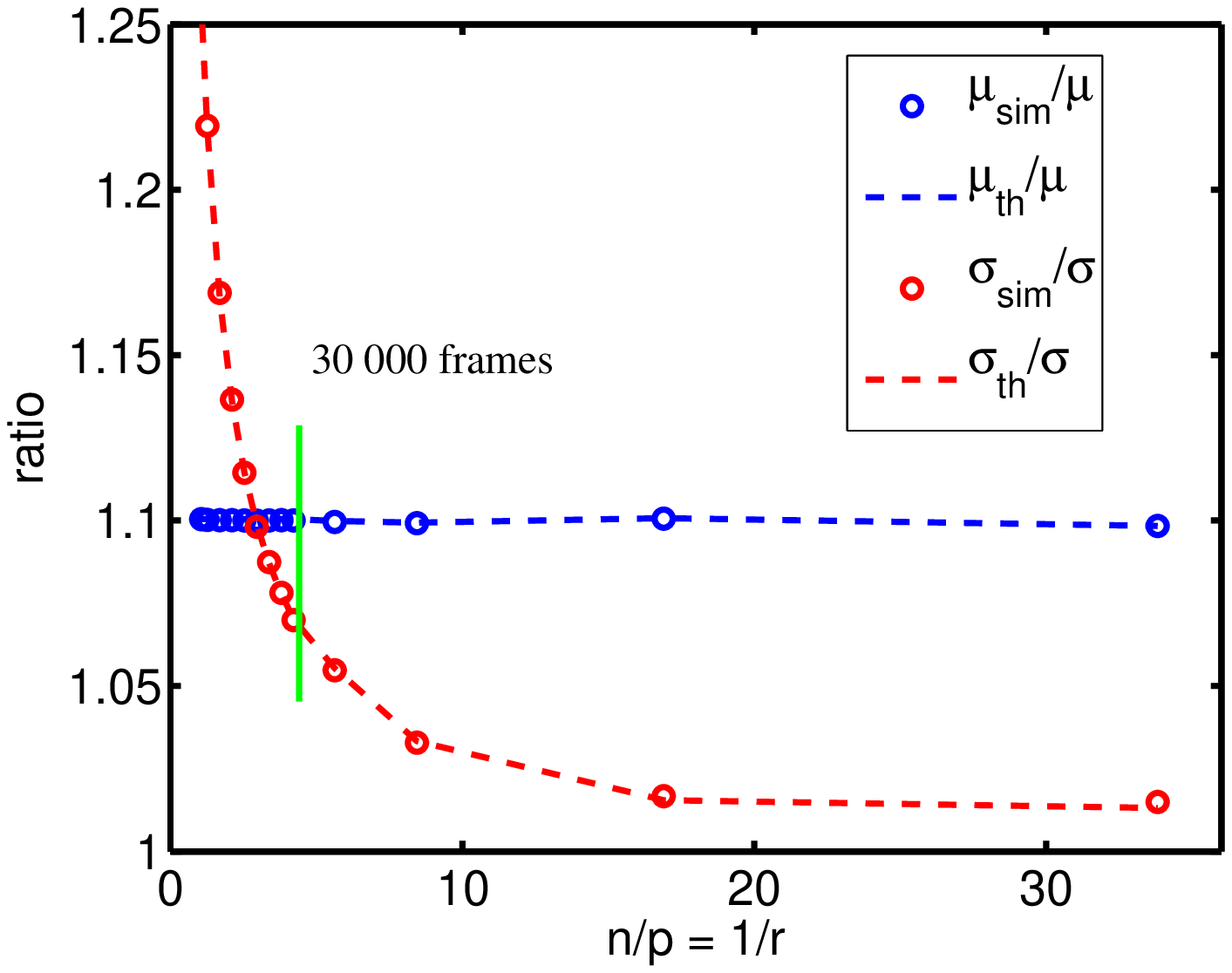}
\includegraphics[width = 0.7\columnwidth,trim = 0mm 0mm 10mm 5mm,
clip]{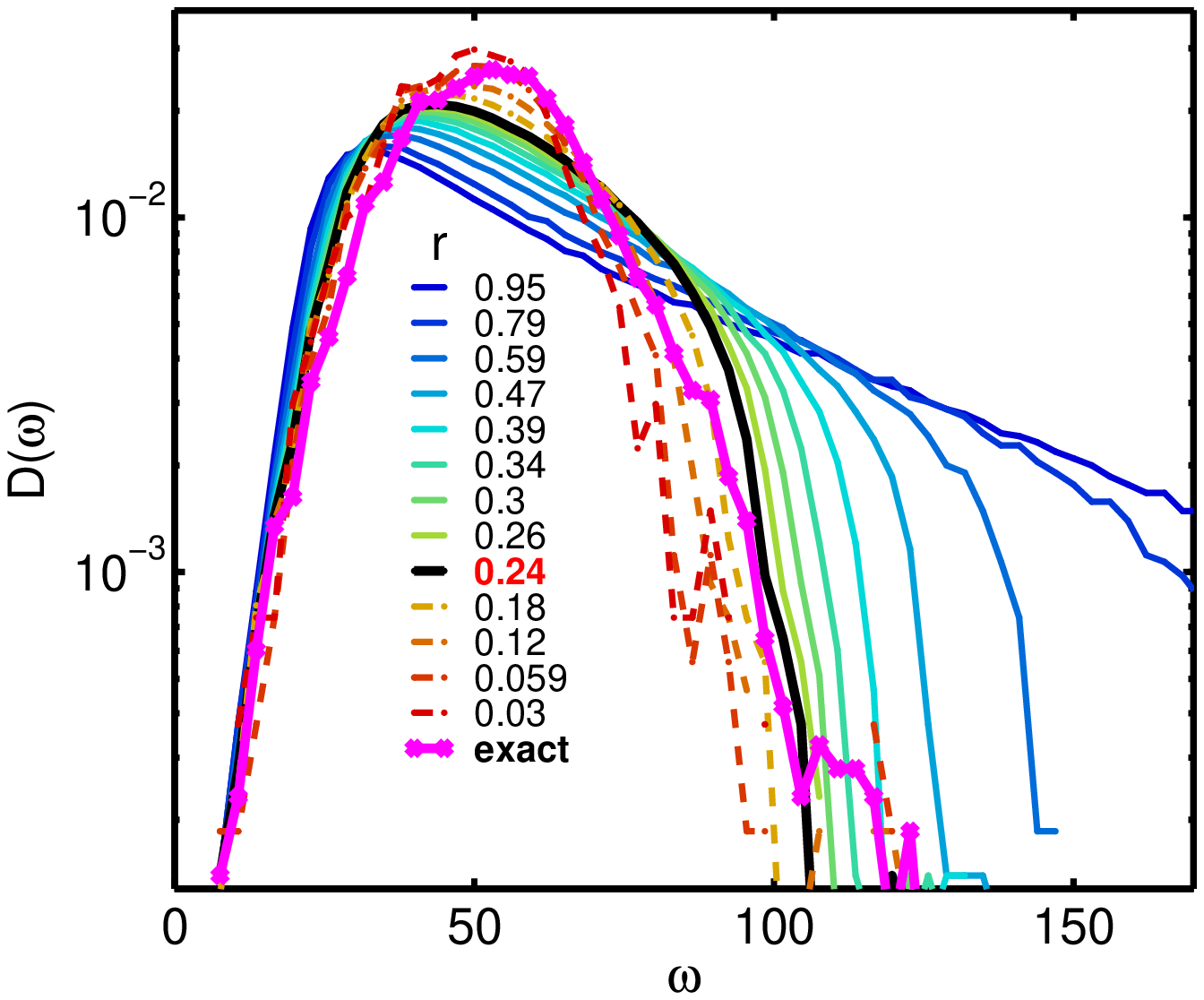}
\includegraphics[width = 0.75\columnwidth,trim = 0mm 0mm 10mm 5mm,
clip]{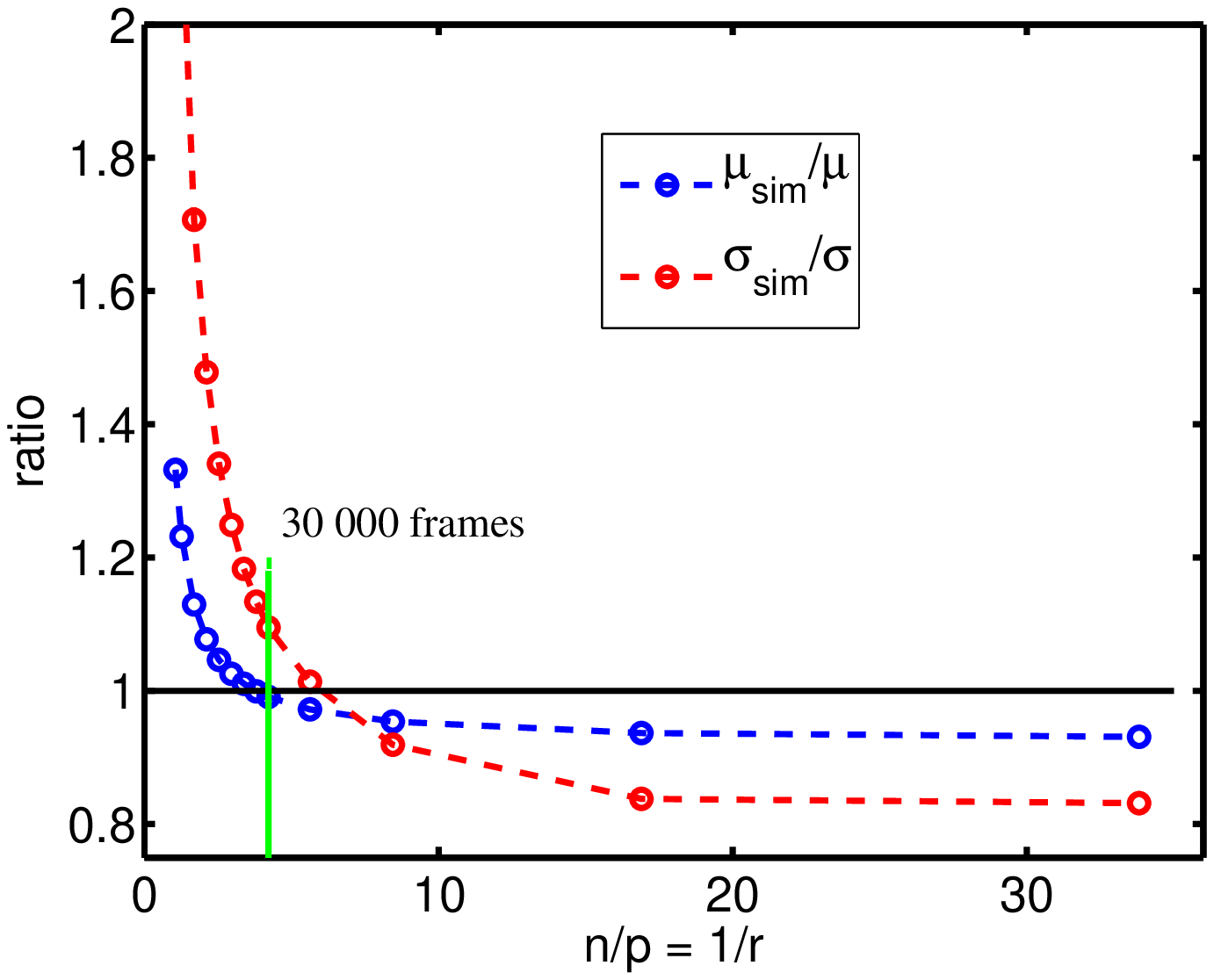}
\caption{Statistical convergence analysis for the NIPA system
\cite{PRL_NIPA_2010}, using the `random vector' model (see text). 
Top left: Numerical eigenvalue distribution for different values of $r$ as
indicated in the legend. The distribution associated with the experimental value
$r=0.24$ is plotted in black and the asymptotic $r\rightarrow 0$ one is plotted
in pink. Bottom left: Same for the density of states. Top right: Numerical (dots, subscript `sim') 
and theoretical (dashed lines, subscript `th') results for the mean and
variance of $P(\lambda)$ normalized by the exact result. The experimental value
of $r=0.24$ is indicated by a green line. Bottom right: Same for the density of
states, numerical result only (dots and dashed lines, subscript `sim'). The green line marked `30 000 frames' 
corresponds to the number of samples in the experimental NIPA system \cite{PRL_NIPA_2010}.}
\label{fig:nipa-statan}
\end{figure*} 

To do so we assume that the applicability conditions of the theorem detailed
in appendix~\ref{appendix_MP} are met. The
finite moments condition corresponds broadly speaking to a long-lived metastable
state with short-time fluctuations of the particles in well-defined cages. The
metastability condition can be tested through the methods of the
section~\ref{test_metastability}. For an equilibrium system, the short-time
fluctuations of the particles in their cages are thermal and can hence well be
approximated as gaussian within the range of linear response. For a
non-equilibrium system such as shaken granular systems, this condition needs to
be verified experimentally. The independence of the different samplings of the
system can be tested through the methods of section~\ref{test_stat}.

If we have now constructed \Cp in these conditions, how does the experimental
resolution affect the properties of the matrix and how does it interplay with
the finiteness of $r$? How does it modify the spectral properties of the true
correlation matrix, the one would obtain in the limits of infinite resolution and
$r\rightarrow 0$?  We model the resolution as an additive Gaussian noise of
variance $\epsilon^{2}$, where $\epsilon$ is the resolution-induced uncertainty of the
particle positions. If $C_0$ is the true correlation matrix, i.e. the one of the
shadow system, one easily shows that for Gaussian fluctuations of the particle
positions, the finite resolution correlation matrix is $\tilde C_0=C_0 +
\epsilon^2 I$, where $I$ is the identity matrix. Then the measured \Cp is $\tilde
C_r$, such that ${\rm lim}_{n\rightarrow\infty}\tilde C_r=\tilde C_0$. The
Mar\u{c}enko Pastur theorem is still valid and one finally relates the
eigenvalues distribution of $\tilde C_r$ and the true correlation matrix $C_0$. 

As for the case without noise, there is no explicit form for the distributions,
but we show in appendix~\ref{appendix_MP} that the mean and variance of the
eigenvalue distribution of \Cp converge with $r\rightarrow 0$ and
$\epsilon/\mu_0 \rightarrow 0$ in the following way:
\begin{align}
 & \tilde{\mu}_{r} = \mu_{0}+\epsilon^{2} \nonumber \\
& \tilde{\sigma}^{2}_{r} = \sigma^{2}_{0} +r \mu_{0}^{2} +r\epsilon^{4},
\label{eq:sigma-lambda0}
\end{align}
where the quantities labeled $0$ refer to the exact eigenvalue distribution and
the ones labeled $r$ to the measured quantities.
Note that for uncorrelated particles $\mu_0$, would be the typical cage
size squared, i.e. the value of the plateau in the mean square displacement
curves, preceding the onset of diffusive behavior. Interestingly,
while the resolution directly impacts the mean, it only changes the width
at $\mathcal{O}(\epsilon^{4})$, and the correction is moreover multiplied by $r$. As
a result the noise dependence of the eigenvalue distribution is in fact much weaker
than one might have expected.

We now present the results of a statistical model specifically developed to test the validity of
the experimental results for colloidal NIPA particles~\cite{PRL_NIPA_2010}. 
In the experiment, $N=3600$ particles forming quasi two dimensional packings deep in the glassy phase
were imaged with a confocal microscope, yielding $n=30000$ independent samples of their positions per run.
In the language of the Mar\u{c}enko-Pastur theorem, this corresponds to $r=0.24$.
The combination of optical resolution and sub-pixel accuracy particle tracking led 
to an estimated uncertainty in the particle positions of $\epsilon=0.007\mu m$.

We build a `random vector' model by constructing a model \Cp where we repeatedly
sample from an uncorrelated multivariate normal distribution coresponding to the
experimentally measured cages sizes, and with added noise at the experimental amplitude.
The covariance matrix of the model is then $\tilde{C}_{ij}=(\lambda_{i}+\epsilon^{2}) \delta_{ij}$, where the $\lambda_{i}$
are the diagonal elements of one of the measured \Cp from ~\cite{PRL_NIPA_2010}.
While this model resembles the `Random Model' introduced in section ~\ref{PCA}
detailing PCA, our aim here is different: we only intend to illustrate equation~\ref{eq:sigma-lambda0}
with a model where we know the exact eigenvalue distribution in the absence of
noise and for $r\rightarrow 0$. By construction, this is just $P(\lambda_{i})$,
the distribution of diagonal elements or cage sizes which enter the model.

Figure \ref{fig:nipa-statan} (top-left) shows the convergence of the numerical
eigenvalue distribution $\tilde{P}_{r}(\lambda)$ as $r\rightarrow 0$ at fixed
noise.  The eigenvalue distribution corresponding to the experimental value
$r=0.24$ is shown in black. It appears to be relatively close to the exact
distribution shown in pink. The numerically obtained first two moments of the distributions
exactly match the theoretical result from equation~(\ref{eq:sigma-lambda0}) (fig.~\ref{fig:nipa-statan}(top-right).
The curves ndicate that for the experimental value of $r$ (green vertical line), the error made compared to
the true converged noiseless distribution is of the order of $10\%$. 

For equilibrium systems, one is mostly interested in the density of states.
Unfortunately, when performing the change of variables
$\omega=\sqrt{k_BT/m\lambda}$, there is no straightforward and reliable way to
calculate the moments of the distribution and we have to rely mostly on
numerical results. Only the asymptotic values in the limit $r\rightarrow 0$
could be obtained (see appendix~\ref{appendix_MP}). In the bottom row of
figure~\ref{fig:nipa-statan} we have applied the transformation $\omega =
1/\sqrt{\lambda}$ numerically to obtain the density of states and its first two
moments for the `random vector’ model. The density of states obtained for the
experimental value of $r=0.24$ again matches pretty well the asymptotic one.
Note that we find a fortitous compensation in the moments of the distributions:
in the limit $r\rightarrow 0$, one can show that a finite resolution
systematically leads to an underestimation of both the mean and the variance.
Since working at finite $r$ always leads to an overestimation of these moments,
it compensates the above underestimation. In the present case,
the experimental value of $r$ (green line) is very close to the point where the
compensation is perfect. Note that this is pure coincidence related to the
specific values of the moments of the true distribution, and that even so it
does not validate the detailed shape of the distribution. 

In the case of the experimental data obtained with NIPA particles~\cite{PRL_NIPA_2010}, and
for which correlations are present, the authors tested convergence of the
density of states based on the methods outlined here and found scaling of mean
and width consistent with our example. The error from experimental and
statistical resolution was estimated in the $10\%$ range at $r=0.24$.

We can make a similar comparison for the hard sphere colloid setup of Ghosh et al.~\cite{PRL_Colloids2010}.
For this experiment, the estimated number of particles is $N=2000$, and approximately $n=4000$ samples at a resolution of $\epsilon=0.03\mu m$ were taken
to obtain the DOS. While the resolution effect is comparable to the NIPA setup, we obtain $r\approx 1$. This introduces a large error into the measurement of the DOS,
similar to the $r=0.95$ curves in the left panels of Figure~\ref{fig:nipa-statan}. We estimate an error of $40\%$ and $200\%$ on the mean and the width of the DOS, respectively.
However, as detailed in the previous section, even at this $r$ it is still possible to obtain the relevant low energy modes~\cite{Ghosh11}.

\subsection{Rattlers}
%*******************************************************************************
\label{rattlers}

Close to jamming, there are few particles, the so called rattlers, which do no
participate to the rigidity of the structure: removing them does not alter
mechanical stability. When computing for instance the average coordination of
the packing, these rattlers  need to be identified and excluded. For simulations
of soft spheres, the procedure is straightforward. The number of contact is
known for each particle and those which have less than two contacts are
considered as rattlers. For simulations of hard spheres just below jamming, the
contacts are defined by computing the transfer of momentum during successive
collisions. Again the particles with less than two contacts are defined as
rattlers, but also those which have a significant smaller momentum transfer
per contact as compared to the average. Such a criterion was proposed 
in~\cite{BritoWyart_JCP2009}, where it was shown that particles with
momentum transfer per contact smaller than $2\%$ of the average value could
safely be considered as rattlers. Varying this threshold in the range
$0.5-5\%$ did not alter the results.

We now offer an alternative way to identify rattlers in experiments by using the
Principal Component Analysis (PCA) itself and compare it to the above method
used for hard spheres. As a matter of fact, the PCA diagonalizes the dynamics
on its dominant features. Since the rattlers are not tightly confined by their
neighbors, they move significantly more than other particles. Also such motion
are by definition localized on a single particle. Hence rattlers correspond to
very singular kind of excitation and thereby tend to jeopardize the modes
structure. Diagonalizing \Cp and analyzing the  modes with high
eigenvalues and  a very high participation ratio is thus a good strategy to
identify rattlers. Here the participation ratio of a given mode $m$  is simply
defined as $P(m)=\sum_{i=1}^N (\delta r_i^m)^4$, where $\delta
r_i^m$ is the displacement of the particle $i$ on the mode $m$. Given that the
modes are normalized, $\sum_{i=1}^N (\delta r_i^m)^2 = 1$ so that a perfectly
delocalized mode has a participation ratio of $1/N$, whereas a mode fully
localized on a single particle would have a participation ratio of order one.

\begin{figure}
%\centering
\includegraphics[width = 0.85\columnwidth, 
clip,angle=-90]{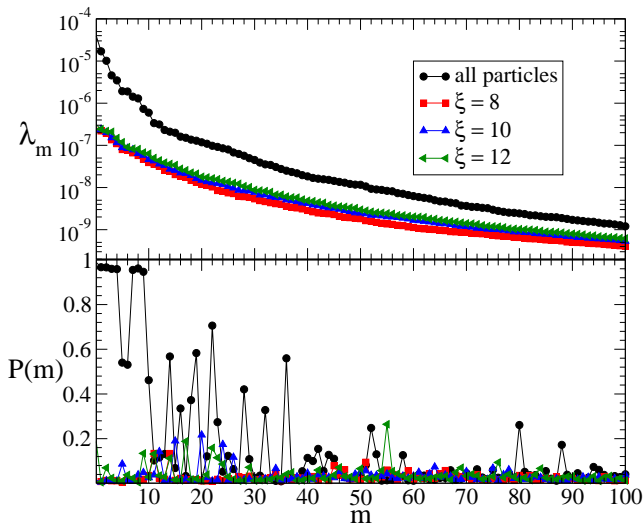}
\caption{Eigenvalues of \Cp and participation ratio as defined in the text for a
system of $N=256$ hard disks below jamming ($\phi = 0.83878$; $\phi_J=0.838865$)
with its rattlers (black curves) and having removed them according to different values
of the thresholding criteria $\xi$ as indicated in legend. See text for the definition of $\xi$.}
\label{fig:ratio_rattlers}
\end{figure}

\begin{figure}
\centering
\includegraphics[width = 0.72\columnwidth, 
clip,angle=-90]{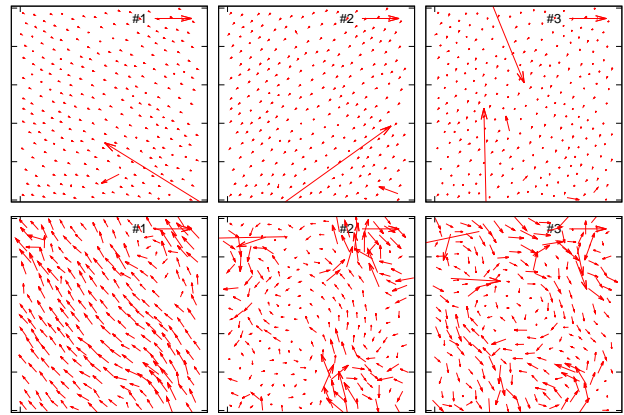}
\caption{Eigen modes of \Cp for a system of $N=256$ hard disks below
jamming ($\phi = 0.83878$; $\phi_J=0.838865$) with its rattlers (top row)
and having removed them (bottom row).}
\label{fig:modes_rattlers}
\end{figure}

The method is illustrated in the case of a system of  $N=256$ hard disks just
below jamming ($\phi = 0.83878$; $\phi_J=0.838865$) within a long lasting
metastable state of total duration $T=4.6\times 10^6$ collisions per particles
with $n=23000$ independent snapshots and $r=0.021$.
Figure~\ref{fig:ratio_rattlers} shows the spectrum and the participation ratio
of the modes of \Cp  computed with and  without rattlers.  One easily identifies
the modes with a simultaneously large eigenvalue and high participation ratio.
Three of these modes are plotted on figure~\ref{fig:modes_rattlers}-top.
Here also, one immediately identifies the rattlers with huge displacements as
compared to the other particles. The rattlers in each modes are then  defined
as the particles having a displacement amplitude larger than $\xi$ times the
average displacement inside the mode: $\delta r_j^{m} \ge \langle \delta r^{m} \rangle + \xi
\sigma _{\delta r}^{m}$, with $\sigma _{\delta r}^{m}$ being the standard
deviations of the displacement on the mode $m$. One must fix a threshold;
however because the modes selected by the procedure concentrate most of
the motion on the rattlers, it is very easy to fix the threshold and the results
are robust to the choice of $\xi$, as it can  verified in the figure~\ref{fig:ratio_rattlers}.
Once the rattlers are identified -- sometime in several modes --, one eliminates
them and recompute \Cp. One then easily checks that the extremely localized
modes have disappeared, together with the rattlers,  figure~\ref{fig:modes_rattlers}-bottom.
Note that most of the particles identified as rattlers using this criterion are 
identical to those identified with the previous criterion used in~\cite{BritoWyart_JCP2009}.
The mismatch between both criteria is of the order of $10\%$ of the rattlers identified.

%*******************************************************************************
\section{Conclusions}
%*******************************************************************************
We have reviewed various situations for which the study of \Cp, the covariance
matrix of the positions of a system of particles can be used to infer dynamical
properties of the system. It has been emphasized that for systems without
equipartition of the energy, or without an independent knowledge of the energy
repartition amongst the modes, the spectrum of \Cp cannot be transposed into the
density of states of the system.  Still, it conveys relevant informations about
the dynamics and physicists would gain in exploring the path already followed.
We have also insist on the physical interpretation of the modes obtained via the
analysis in terms of the vibrational properties of the ‘shadow system’, which
must not be confused with the experimental system, even an idealized one.

We then have seen the numerous steps one has to go through in implementing the
analysis. Metastability of the reference state, harmonicity of the dynamics,
statistical independence, convergence and experimental resolutions are all key
elements to care of. How to deal with rattlers was finally discussed. Following
the whole procedure requires as always a bit of physical insights, a good
knowledge of the system of interest and patience. We have \emph{not} discussed
the new physics one can finally extract from the analysis. It certainly depends
on the specificity of each system, but in any case one should remember that the
analysis remains linear in essence.

We hope that in this light the present review of the methods associated to the
correlation matrix approach will be useful to the community.

%*******************************************************************************
\section{Acknowledgements}
%*******************************************************************************
We would like to thank J.-P. Bouchaud and G. Biroli for inspiring the
correlation matrix approach and pointing to the Mar\u{c}enko-Pastur theorem and
acknowledge the soft matter groups at U Penn for collaborating and sharing their
data on the NIPA system. None of the present work would have been done without
the numerous discussions, we have had at various level with those involved in
the analysis in terms of vibrational modes of experimental and numerical data. 
In particular we thank Wouter Ellenbroek, Andrea Liu and Corey O'Hern for
helpful discussions on the interpretation of modes. A special thought to Wim van
Sarloos and Martin Van Hecke who hosted us in Leiden and always encouraged us to
persevere in this delicate game, the rules and traps of which we have tried to
expose in this paper. 

%*******************************************************************************
\section{Appendix}
%*******************************************************************************

\subsection{Newtonian dynamics}
%*******************************************************************************
\label{appendix_Newton}

Starting from the Newton equations linearized around the reference state
$|r^0\rangle$, 
\begin{equation}
 |\delta  \ddot{r} \rangle + {\bf D} |\delta  {r} \rangle = 0,
\label{newtonian}
\end{equation}  
\noindent
where  ${\bf D}={\bf K}/m$ is called the {\it dynamical matrix}, the eigenmodes
of ${\bf D}$, also called the {\it vibrational modes} of the system are defined
by ${\bf D} |\lambda_q\rangle = \omega_q^2 |\lambda_q\rangle$, where  $\omega_q$
are the vibrational frequencies.
The solution of  Eq.(\ref{newtonian}) is given by:
\begin{equation}
|\delta  {r} \rangle = e^{-i\sqrt{\bf D}t}  |\delta  {r(0)} \rangle,
\end{equation}
where  $|\delta  {r(0)} \rangle$ is the displacement field at time zero. 
Replacing this solution in Eq.(\ref{Cp}), one has
\begin{equation}
C_p = \left< ~  e^{i\sqrt{\bf D} t} |\delta r(0) \rangle  \langle \delta r(0)|
e^{i\sqrt{\bf D} t}   \right>;
\end{equation}
then writing  $|\delta {r(0)} \rangle$ in the eigenbasis of \D,
$\{|\lambda_q\rangle\}$:
\begin{equation}
|\delta {r(0)} \rangle = \sum_{q} \alpha_q | \lambda_q \rangle,
\end{equation}
where $\alpha_q =  \langle  \lambda_q |\delta r(0) \rangle$, is the amplitude of
the initial condition on the mode $|\lambda_q\rangle$, one obtains
\begin{eqnarray}
{\bf C_p} = \left< ~ \sum_{q,k} \alpha_q \alpha_k^*e^{-i (\omega_q-\omega_k) t}
| \lambda_q \rangle  \langle \lambda_k| ~\right>
\label{Cp_avecs}
\end{eqnarray}
Provided that the time average $\langle . \rangle$ is performed on a large
enough interval, large as compared to the inverse of the minimal gap between
adjacent frequencies $(\omega_q-\omega_m)$, one has 
\begin{eqnarray}
\left< ~ e^{-i (\omega_q-\omega_m) t} \right> = \delta_{q,m}, 
\label{average1}
\end{eqnarray}
and
\begin{equation}
{\bf C_p} = \sum_{k} \alpha_k^2 | \lambda_k \rangle  \langle \lambda_k|.
\end{equation}
Using the orthogonality of the eigenvectors  $ \{ | \lambda_q \rangle \}$,
$\langle \lambda_k| \lambda_m \rangle = \delta_{k,m}$, one finally obtains the
eigenvalue equation of \Cp:
\begin{eqnarray}
{\bf C_p} | \lambda_q \rangle =  \alpha_q^2| \lambda_q \rangle.
\end{eqnarray}

\noindent 
At equilibrium, the initial condition is thermalized and the energy is equally
distributed among the modes. Each mode of frequency $\omega_q$ and amplitude
$\alpha_q$ carries an energy $m\alpha_q^2 \omega_q^2/2=k_BT/2$.  Hence the
eigenvalues $\lambda_q$ of \Cp and the vibrational frequencies $\omega_q$ of the
dynamical matrix are related through
\begin{equation}
 \lambda_q = \alpha_q^2 = \frac{k_BT}{m\omega_q^2}.
\label{evalues_Newton}
\end{equation}
\subsection{Overdamped  Langevin dynamics}
%*******************************************************************************
\label{appendix_Langevin}
Starting from the Langevin equation for an overdamped linearized dynamics around
the reference state $|r^0\rangle$:
\begin{equation}
|\delta \dot{r}(t)\rangle =  - \frac{{\bf K}}{\mu} |\delta{r}\rangle +
\frac{1}{\mu}|\eta(t)\rangle, 
\label{Langevin}
\end{equation}
where $\mu$ is the viscous damping and $|\eta(t)\rangle$ is a white noise of
amplitude $\Gamma$: $\langle \eta(t')  |\eta(t'')\rangle = \Gamma \delta(t'-t'')
$, one introduces the operator  ${\cal  L} = {\bf K}/\mu$ whose eigenvalue
equation is :  ${\cal L}|\lambda_q\rangle = m \omega_q^2 /\mu
|\lambda_q\rangle$. $\tau_q =\mu/\kappa_q=\mu/(m\omega_q^2)$ is the relaxation time of the
system along the eigenmode $|\lambda_q\rangle$.\\

\noindent
The solution of the Eq.(\ref{Langevin}) can be written as
\begin{equation}
 |\delta{r}\rangle = e^{-{\cal L} t} |\delta  {r(0)} \rangle + 
\frac{1}{\mu}\int_0^{t}e^{-{\cal L} (t-t')}|\eta(t')\rangle dt' 
\label{solution_Lanvegin}
\end{equation}

\noindent
Replacing this solution \ref{solution_Lanvegin} in  Eq.(\ref{Cp})  one obtains 
an expression with four terms, only two of which are not zero because the  noise
$| \eta(t)\rangle$ is assumed not to be correlated with the position field
$|\delta  {r(0)} \rangle$. One ends up with:

\begin{eqnarray}
\lefteqn{C_p = \left< ~  e^{-{\cal L}t} |\delta r(0) \rangle  
          \langle \delta r(0)| e^{-{\cal L}t}   \right> +} \\ \nonumber 
&&\frac{1}{\mu^2}\left< ~ \int_0^t dt'' \int_0^t dt' e^{-{\cal L}(t-t')} |
\eta(t')\rangle  
\langle \eta(t'')| e^{-{\cal L}(t-t'')}   \right>
\label{Cp_lambda}
\end{eqnarray}

\noindent
Expressing  $|\delta {r(0)} \rangle$  and $|\eta(t)\rangle$ in the eigenbasis
$\{|\lambda_q\rangle\}$ of \LL
$|\delta r(0) \rangle = \sum_{q} \alpha_q |\lambda_q \rangle$ and 
$|\eta(t)\rangle= \sum_{q} \beta_q(t) |\lambda_q \rangle$,  with $\alpha_q =
\langle \lambda_q |\delta r(0) \rangle $ and $\beta_q = \langle \lambda_q
|\eta(t)\rangle$, the first term of the previous equation, which we note $T_1$
turns into:

\begin{eqnarray*}
T_1 &=& 
\left< ~   \sum_{q,k} e^{-(1/\tau_q+1/\tau_k) t} 
\alpha_q\alpha_k^*|\lambda_q\rangle   \langle \lambda_k|   \right> \\
&=&  \sum_{k} e^{-2t/\tau_k}  \alpha_k^2  |\lambda_k\rangle   \langle
\lambda_k|,
\end{eqnarray*}

\noindent 
where the last equality is justified by the fact that the components on each
mode of the initial conditions are uncorrelated
$\left< ~ \alpha_q\alpha_k^*  \right> = \alpha_q\alpha_k^* \delta_{k,q}$.

We now turn our attention to the second term in the Eq.(\ref{Cp_lambda}),
referred to as $T_2$:
\begin{widetext}
\begin{eqnarray*}
T_2 &=& \frac{1}{\mu^2}\left< \int_0^t dt''\int_0^t dt' \sum_{q,k}  
e^{-1/\tau_q (t-t')} \beta_q(t') |\lambda_q \rangle \langle
\lambda_k|\beta_k(t'')
e^{-1/\tau_k(t-t'')} \right>\\
&=&  \int_0^t dt'' \int_0^t dt' \sum_{q,k}  
e^{-(t-t')/\tau_q-(t-t'')/\tau_k}
\left< ~  \beta_q(t')  \beta_k(t'') \right> |\lambda_q \rangle 
\langle \lambda_k|    \\
\end{eqnarray*}
\end{widetext}

\noindent
Now assuming that the components of the noise on the modes are also uncorrelated
 $\left< ~ \beta_q(t') \beta_k(t'')\right> = \Gamma  \delta_{q,k} 
\delta(t'-t'')$ one obtains for $T_2$:

\begin{eqnarray*}
T_2 &=& \frac{\Gamma}{\mu^2} \sum_{k} \int_0^t dt' e^{-2(t-t')/\tau_k}|\lambda_k
\rangle \langle \lambda_k| \\
  &=& \frac{\Gamma}{\mu^2} \sum_{k}\frac{\tau_k}{2}(1-e^{-2 t/\tau_k})|\lambda_k
\rangle \langle \lambda_k| 
\end{eqnarray*}

Finally applying  $C_p$ on the eigenvector $ |\lambda_q \rangle$ one obtains the
eigenvalue equation for \Cp:

\begin{eqnarray}
C_p |\lambda_q \rangle =  
 \left[ \left( \alpha_q^2 - \frac{\Gamma \tau_q}{2\mu^ 2}\right) 
e^{-2t/\tau_q} + \frac{\Gamma \tau_q }{2\mu^ 2}\right] |\lambda_q \rangle
\end{eqnarray}

\subsection{Out of Equilibrium Stochastic Forcing}
%*******************************************************************************
\label{appendix_stochmech}
To generalize on the two equilibrium cases of Newtonian dynamics and Langevin
dynamics, we now consider a non-equilibrium system with both inertia and damping
and a colored noise spectrum. We choose to stick to a Langevin type description
of the dynamics, where  the damping is linear and single particle, as
appropriate for particles in a newtonian fluid bath, but not necessarily for a
large scale mechanical excitation. The generic equations of motion will then be:

\begin{equation}
 m|\delta \ddot{r}(t) \rangle + \mu |\delta \dot{r}(t) \rangle = -{\bf K} |
\delta r(t) \rangle + |\eta(t)\rangle
\end{equation}

We now expand in the modes of $K$, using the same notation as for the Langevin
case, ${\bf K} |\lambda_{q} \rangle = \kappa_{q} |\lambda_{q} \rangle$, and $ |\delta r (t)\rangle=\sum_{q} \alpha_{q}(t) |\lambda_{q}\rangle$  . We
define the colored noise in the basis of the modes as
\begin{align}
& |\eta(t)\rangle=\sum_{q} \eta_{q}(t) |\lambda_{q}\rangle \quad \nonumber \\
& \text{with} \quad \langle \eta_{q}(t) \eta_{k} (t') \rangle = \Gamma_{q}
\delta_{q k} \delta(t-t'). 
\end{align}
\noindent
In the eigenbasis of the modes, the equations of motion are:
\be \ddot{\alpha}_{q}(t) + \frac{\mu}{m} \dot{\alpha}_{q}(t) +
\frac{\kappa_q}{m} \alpha_{q}(t) = \eta_{q}(t). \ee

% done
% {\color{blue}{\bf   Where is ${\alpha}$  defined ??? }}.

We first solve the homogeneous equation, in the absence of noise. The solutions
are either oscillatory or overdamped depending on the relative importance of
inertia and damping.

For $(\mu/m)^2 < 4 \kappa_q /m$, i.e. the low damping limit, we find two
oscillating solutions
\begin{align}
& \alpha_{q}^{s}(t) = \alpha_{q}^{s}(0) e^{-\frac{\mu t}{2 m}} \sin \left[
\frac{1}{2} \tilde{\Omega}_{q} t \right] \nonumber \\
& \alpha_{q}^{c}(t) = \alpha_{q}^{c}(0) e^{-\frac{\mu t}{2 m}} \cos \left[
\frac{1}{2} \tilde{\Omega}_{q} t \right] \nonumber \\
& \tilde{\Omega}_{q}=\sqrt{4 \frac{\kappa_q}{m} - \left(\frac{\mu}{m}\right)^{2}
}
\end{align}
Clearly, the limit $\mu/m \rightarrow 0$ corresponds to newtonian dynamics.

 For $(\mu/m)^2 > 4 \kappa_q /m$, i.e. the strong damping limit, we obtain two
decaying solutions
\begin{align}
& \alpha_{q}^{\pm}(t)=\alpha_{q}^{\pm}(0) \exp \left \lbrace \frac{1}{2} \left[
\frac{-\mu}{m} \pm \Omega_{q} \right] t \right \rbrace \nonumber \\
& \Omega_{q}=\sqrt{ \left(\frac{\mu}{m}\right)^{2}-4 \frac{\kappa_q}{m}}
\end{align}
It can be shown that in the limit $m/\mu \rightarrow 0$,  the `$+$'-solution
corresponds to the Langevin homogeneous solution $\alpha_{q}(t) = \alpha_{q}^{+}
(0) e^{-\kappa_q t/\mu}$, while the `$-$'-solution decays infinitely fast. 

To solve the inhomogeneous equation, we find a particular solution
$\alpha_{q}^{p}(t)$ through the method of variation of constants based on the
homogeneous solutions.
In the oscillatory regime, we find
\begin{align}
&\alpha_{q}^{p}(t)=\int_{0}^{t} \frac{\eta_{q}(t')}{m \tilde{\Omega}_{q}}
e^{\frac{\mu}{2 m} (t-t')} 2 \sin \left( \frac{\tilde{\Omega}_{q}}{2}
(t-t')\right)
\end{align}
while for the damped regime, the particular solution is given by:
\begin{align}
&\alpha_{q}^{p}(t)=\int_{0}^{t} dt' \frac{\eta_{q}(t')}{m \Omega_{q}}
e^{\frac{\mu}{2 m} (t-t')} 2 \sinh \left( \frac{\Omega_{q}}{2} (t-t')\right)
\end{align}
Both solutions are similar in spirit to the particular solution of the Langevin
equation: the source of continued motion in the system is the colored noise, and
in the long time limit, \emph{only} the motion due to the noise remains. From
this point of view, the memory a non-dissipative system such as the Newtonian
case retains of the initial conditions is singular.

We can now calculate $C_{p}$ by performing an ensemble average over initial
conditions and the noise. In particular, we assume that we can replace the noise
correlations by their expectation value $\langle \eta_{q}(t) \eta_{k} (t')
\rangle = \Gamma_{q} \delta_{q k} \delta(t-t')$. We also assume that noise and
initial conditions do not cross-correlate, and that in ensemble-average, the
initial conditions of different modes are independent of each other. 
Then only the diagonal terms in the modes remain and, schematically, $C_{p}$ is
given by:
\begin{align}
& C_{p}(t) = \sum_{q} \langle \alpha_{q}^{1}(t)^{2} \rangle + \sum_{q} \langle
\alpha_{q}^{2}(t)^{2} \rangle + \sum_{q} \langle \alpha_{q}^{p}(t)^{2} \rangle
\end{align}

For the oscillatory case, we obtain
\begin{align} 
&C_{p}= \sum_{q} e^{-\mu t/m} \left[ \langle (\alpha_{q}^{s}(0))^{2} \rangle
\sin^{2}(\tilde{\Omega}_{q} t/2) \right. \nonumber \\
 & + \left. \langle (\alpha_{q}^{c}(0))^{2} \rangle \cos^{2}(\tilde{\Omega}_{q} t/2) \right]
+ \frac{\Gamma_{q}}{2 \mu \kappa_q} - \frac{2 \Gamma_{q} e^{-\mu t/m}}{m \mu
\tilde{\Omega}_{q}^{2}} \nonumber \\
& - \frac{\mu}{m} \frac{\Gamma_{q} e^{-\mu t/m}}{2 \mu \kappa_q
\tilde{\Omega}_{q}^{2}} \left[ \tilde{\Omega}_{q} \sin \left(\tilde{\Omega}_{q}
t \right) - \frac{\mu}{m} \cos \left( \tilde{\Omega}_{q} t \right) \right]
|\lambda_{q} \rangle \langle \lambda_{q} |
\label{eq:ev1}
\end{align}

For the damped case, we find
\begin{align} 
&C_{p} =  \sum_{q} e^{-\mu t/m} \left[ \langle (\alpha_{q}^{+}(0))^{2} \rangle
e^{\Omega_{q}t} +\langle (\alpha_{q}^{-}(0))^{2} \rangle e^{-\Omega_{q}t}
\right]  \nonumber \\
&+ \frac{\Gamma_{q}}{2 \mu \kappa_q} \left(1+e^{-\mu t/m} \right) 
 - \frac{\mu}{m} \frac{\Gamma_{q} e^{-\mu t/m}}{4 \mu \kappa_q \Omega_{q}^{2}}
\left[ \Omega_{q} \left(e^{\Omega_{q}t}-e^{-\Omega_{q}t} \right) \right. \nonumber \\
& + \left. \frac{\mu}{m}
\left(e^{\Omega_{q}t}+e^{-\Omega_{q}t} -2 \right) \right] |\lambda_{q} \rangle
\langle \lambda_{q} | 
\label{eq:ev2}
\end{align}
which in the over-damped limit $m/\mu \rightarrow 0$, defaults to the result for
Langevin dynamics if we assume equilibrium noise $\Gamma_{q} = \Gamma$, $\forall
q$.

In both cases, in the long time limit we recover the an expression similar to
that of the equilibrium result 
\begin{equation}
 C_{p}|\lambda_{q} \rangle = \frac{\Gamma_{q}}{2 \mu \kappa_q} |\lambda_{q}
\rangle
\end{equation}
\noindent where the initial conditions cease to matter.

\subsection{Sampling, Resolution and the Mar\u{c}enko-Pastur theorem}
%**************************************************************************
\label{appendix_MP}

We derive relations which characterize the deviation induced by finite sampling
\emph{and finite resolution} on the spectral properties of \Cp. For a system of
$N$ particles the positions of which are sampled independently $n$ times with a
resolution $\epsilon$, we will establish an adapted form of the
Mar\u{c}enko-Pastur theorem, which then allows to derive relations between the
moments of the experimentally measured eigenvalue distribution and the moments
of the eigenvalue distribution of the true correlation matrix. 
The sampling number $r = Nd/n$,  is defined as the total number of degrees of
freedom $(Nd)^{2}$ divided by the number of measurements $n\times Nd$. 

In its most general form, the Mar\u{c}enko-Pastur
theorem~\cite{MarcenkoPastur,Silverstein,ElKaroui,Burda} is valid on the
ensemble of Wishart random matrices, that is matrices constructed as
$C_{r}=X^{*}X$, where $X$ is a $n\times p$ matrix which can be written as
$X=YC_0^{1/2}$ where the elements of the $n \times p$ matrix $Y$ are identically
independently distributed (i.i.d), with mean 0, variance 1 and a finite fourth
moment and $C_0$ is a $p\times p$ positive definite matrix.
In the following, we shall use the slightly more specialized derivation for
Gaussian distributions found in Burda et al.~\cite{Burda}. We thus assume that
$C_{r}$ is the average of $n$ samplings of $x_{i} x_{j}$, where the variables
$\lbrace x_{i} \rbrace,$ $i=1...p$ are  identically and independently
distributed according to a multivariate normal distribution:
\begin{equation} P(x_{1}, ... x_{p}) = \left [ (2\pi)^p
\text{det}(C_0)\right]^{-1/2} \exp \left [ -\frac{1}{2} x_{i} C_{0 ij}^{-1}
x_{j} \right].
\label{eq:cijdist} 
\end{equation}
In the limit $n\rightarrow\infty$, at fixed $p$, one then clearly has from
equation~(\ref{eq:cijdist}) $\lim_{r\rightarrow 0} C_{r} = C_{0}$. 

For a given correlation matrix $C$, define the (negative) moment generating
function of the eigenvalue spectrum $P(\lambda)$
\begin{equation} m(z) = \sum_{k=1}^{\infty} \frac{m_{k}}{z^{k}},
\label{eq:momdef} \end{equation}
where $m_{k} = \langle \lambda^{k} \rangle$. Also define the Stieltjes transform
of the eigenvalue distribution $P(\lambda)$ written as a sum over the poles of
$(C-z I)^{-1}$ at the eigenvalues
\begin{equation} s(z)=\frac{1}{p} \text{Tr}\left( (C-z I)^{-1}\right), \quad z
\in \mathfrak{C}^{+}. \end{equation}
Both are related by 
\begin{equation} m(z) = - z s(z)-1. \label{eq:moment} \end{equation}
To see this, note that in the eigenbasis of $C$, we can write the Laurent series
of the right hand side as
\begin{equation} \sum_{k=1}^{\infty}  \frac{1}{p} \sum_{l=1}^{p}
\frac{\lambda^{k}_{l}}{z^{k}}. \end{equation}

Let us denote $P_{0}(\lambda)$ , $m_{0}(z)$, $s_{0}(z)$ the eigenvalue spectrum,
the generating function and the Stieljes transform of the true correlation
matrix $C_{0}$,  and  $P_{r}(\lambda)$, $m_{r}(z)$, $s_{r}(z)$ the corresponding
quantities for the experimental $C_{r}$. Then the Mar\u{c}enko Pastur theorem
stated in Burda et al.~\cite{Burda} provides an explicit conformal
transformation relating $m_{r}(z)$ to $m_{0}(z)$:
\begin{equation} 
m_{r}(z_{r}) = m_{0}(z_{0}) \quad \text{where} \quad z_{0} = \frac{z_{r}}{1+r
m_{r} (z_{r})}.
\label{eq:confrel}
\end{equation}
Burda et al. then derive relations between the moments $m_{k}^{r}$ and
$m_{k}^{0}$ through writing the Laurent series in $z_{r}$ on both sides, and
then equating the coefficients of the powers of $z_{r}^{-k}$. 

We can adapt this approach to include an experimental gaussian noise term as
follows. If $C_{0}$ is the true correlation matrix, 
let $y_{i} = \delta x_{i} +\xi_{i}$ be the particle fluctuations with the
resolution-induced noise included, so that we have $P(\xi_{i}) = [2\pi
\epsilon^{2} ]^{-1/2} \exp(-\xi^{2}/\epsilon^{2})$.
The probability distribution of a sum of two random variables is obtained
through a convolution of their respective probability distributions, so that we
have
\begin{equation} P(y_{i}, ... ,y_{p}) = \prod_{i=1}^{p} \int d\xi_{i} P(\xi_{i})
P(y_{1}-\xi_{1}, ..., y_{p}-\xi_{p}). \end{equation}
This integral can be calculated by completing the square in the eigenbasis of
$C_{0}$, and the resulting distribution is simply Gaussian 
with variance matrix $\tilde{C}_{0}=C_{0}+\epsilon^{2} I$.

Then the correlation matrix which determines the probability distribution in our
system is $\tilde{C}_{0} = C_{0} +\epsilon^{2}I$ and what is measured is
$\tilde{C}_{r}$, such that $\lim_{n\rightarrow \infty} \tilde{C}_{r} =
\tilde{C}_{0}$. 
The Mar\u{c}enko-Pastur theorem is still valid and we have $\tilde{m}_{r}(z_{r})
= \tilde{m}_{0}(z_{0})$. We can then relate the modified moment generating
function $\tilde{m}_{0}(z_{0})$ to the real moment generating function
${m}_{0}(\tilde{z}_{0})$ through a change of variables $\tilde{z}_{0} =
z_{0}-\epsilon^{2}$. The Stieltjes transform of the new eigenvalue distribution
becomes:
\begin{equation} 
\tilde{s}(z) = \frac{1}{p}\text{Tr}(\tilde{C}-zI)^{-1} = \frac{1}{p}\text{Tr}
(C-\tilde{z}I)^{-1} = s(\tilde{z}). 
\end{equation}
We can then derive the moment generating function:
\begin{equation}
\tilde{m}(z) = -\left(z \tilde{s}(z)\right) -1 = -(\tilde{z} s(\tilde{z})) -1
-\epsilon^{2} s(\tilde{z}).
\end{equation}
The first two terms are nothing but $m(\tilde{z})$, while one can show through a
Laurent expansion that the last term 
is given by $\frac{\epsilon^{2}}{\tilde{z}} [ 1+ m(\tilde{z})]$. The relation
between the moment generating functions becomes then
\begin{equation} \tilde{m}(z) =
m(\tilde{z})\left[1+\frac{\epsilon^{2}}{\tilde{z}}\right]+\frac{\epsilon^{2}}{
\tilde{z}}, \label{eq:mtilde-m} \end{equation}
from which we derive an adapted form of the Mar\u{c}enko-Pastur theorem:
\begin{align} &\tilde{m}_{r}(z_{r})
=m_{0}(\tilde{z}_{0})\left[1+\frac{\epsilon^{2}}{\tilde{z}_{0}}\right]+\frac{
\epsilon^{2}}{\tilde{z}_{0}}, \\ \nonumber 
& \text{where} \quad \:\:\:\:\: \tilde{z}_{0} = \frac{z_{r}}{1+r m_{r} (z_{r})}
- \epsilon^{2}
\end{align}
Finally, this allows us to derive relations between the moments
$\tilde{m}^{k}_{r}$ of the experimentally measured eigenvalue distribution and
the moments of the eigenvalue distribution of the shadow system $m^{k}_{0}$. 
This can be done through a Laurent expansion in $z_{r}$ on both sides, together
with a Taylor expansion in $\epsilon^{2}$. After a considerable amount of
algebra we find to second order in $\epsilon^{2}$:
\begin{align} 
 & \tilde{m}^{1}_{r} = m^{1}_{0} + \epsilon^{2} \nonumber \\
& \tilde{m}^{2}_{r} = m^{2}_{0} + 2 m^1_{0} \epsilon^{2} + r (m^{1}_{0})^{2} +
\epsilon^{4}
\end{align}
The central moments, i.e. the mean $\mu$ and the variance $\sigma^{2}$, are more
convenient and we find 
\begin{align}
 & \tilde{\mu}_{r} = \mu_{0}+\epsilon^{2} \nonumber \\
& \tilde{\sigma}^{2}_{r} = \sigma^{2}_{0} +r \mu_{0}^{2} \label{eq:sigma-lambda}
\end{align}
To this order, the second moment does not depend on the noise, however it comes
in at higher orders. For a system with pure noise equation \ref{eq:sigma-lambda}
gives $\mu_{n} = \epsilon^{2}$ and $\sigma_{n}^{2} = r \epsilon^{4}$ (note that
for $r\rightarrow 0$, the second moment of the noise eigenvalue distribution
vanishes as it should).  Since for large enough $\epsilon$, this has to be
consistent with (\ref{eq:sigma-lambda}), we obtain:
\begin{align}
 & \tilde{\mu}_{r} = \mu_{0}+\epsilon^{2} \nonumber \\
& \tilde{\sigma}^{2}_{r} = \sigma^{2}_{0} +r \mu_{0}^{2} + r
\epsilon^{4}\label{eq:sigma-lambdatot}
\end{align}

To convert these relations into useful relations for the density of states one
needs to perform the change of variable 
$\omega \propto \lambda^{-1/2}$. Unfortunately, we could not find a
straightforward and reliable way to calculate the moments of this distribution
and we had to rely mostly on numerical results. However in the limit
$r\rightarrow 0$, the relation 
$\tilde{C}_{0}=C_{0}+\epsilon^{2} I$ directly translates to the eigenvalues,
i.e. $\tilde{\lambda}_{j}^{0} = \lambda_{j}^{0} +\epsilon^{2}$, 
and we recover equation \ref{eq:sigma-lambdatot} for $r=0$. Here we can perform
the change of variables explicitly, 
and we find to order $\epsilon^{2}$ (written in the most convenient mixture of
direct and central moments):
\begin{align}
 &\tilde{\mu}_{0}^{\omega} = \mu_{0}^{\omega} - \frac{\epsilon^{2}}{2} \langle
\omega^{3} \rangle_{0} \nonumber \\
&\left(\tilde{\sigma}^{2}_{0}\right)^{\omega} =
\left(\sigma^{2}_{0}\right)^{\omega}- \epsilon^{2}\left(\langle \omega^{4}
\rangle_{0} - \langle \omega^{3} \rangle_{0} \langle \omega \rangle_{0} \right)
\label{eq:sigma-omega}
\end{align}
It can be shown that the factor multiplying $\epsilon^2$ in the equation for
$\sigma^{2}$ is strictly positive, 
so that in the limit $r\rightarrow 0$, both mean and variance are \emph{reduced}
proportional to $\epsilon^{2}$.

\bibliography{modes_references}

\end{document}